# Scalar Clouds and Quasinormal Modes

**Gülnihal Tokgöz**

Submitted to the
Institute of Graduate Studies and Research
in partial fulfillment of the requirements for the degree of

Doctor of Philosophy
in
Physics

Eastern Mediterranean University
January 2019
Gazimağusa, North Cyprus

Approval of the Institute of Graduate Studies and Research

                                                      Assoc. Prof. Dr. Ali Hakan Ulusoy
                                                                   Acting Director

I certify that this thesis satisfies all the requirements as a thesis for the degree of Doctor of Philosophy in Physics.

                                                                    Prof. Dr. İzzet Sakallı
                                                   Chair, Department of Physics

We certify that we have read this thesis and that in our opinion it is fully adequate in scope and quality as a thesis for the degree of Doctor of Philosophy in Physics.

                                                                    Prof. Dr. İzzet Sakallı
                                                                           Supervisor

                                                                                    Examining Committee

1. Prof. Dr. Muzaffer Adak

2. Prof. Dr. İzzet Sakallı

3. Prof. Dr. Nuri Ünal

4. Assoc. Prof. Dr. Mustafa Gazi

5. Assoc. Prof. Dr. S. Habib Mazharimousavi

# ABSTRACT

In this thesis, Maggiore's method (MM), which evaluates the transition frequency that appears in the adiabatic invariant from the highly damped quasinormal mode frequencies (QNMFs), is used to investigate the entropy and area spectra of the Garfinkle-Horowitz-Strominger black hole (GHSBH), the four-dimensional Lifshitz black hole with zero dynamical exponent (ZZLBH), and the rotating linear dilaton black hole (RLDBH), respectively. For studying the GHSBH's quantization, instead of ordinary quasinormal modes (QNMs), the boxed QNMs (BQNMs) that are the characteristic resonance spectra of the confined scalar fields in the GHSBH geometry are computed. To this end, it is assumed that the GHSBH has a confining mirror placed in the vicinity of its event horizon. It is then shown how the complex resonance frequencies of this caged black hole (BH) are computed considering the scalar perturbations around the event horizon. Additionally, the resonance modes (RMs) and the RM frequencies (RMFs) of the GHSBH are computed by using the confined scalar waves with high azimuthal quantum number. The entropy and area quantizations are shown to be independent of the GHSBH's parameters, however, both spectra are equally spaced.

The spectroscopy of ZZLBH is studied by solving the Klein-Gordon equation (KGE). Based on the exact solution to the near-horizon (NH) Schrödinger-like equation (SLE) of the massive scalar waves, the QNMs of the ZZLBH are computed by employing the adiabatic invariant quantity. The Dirac equations for the fermionic perturbations of this spacetime are solved in the framework of Newman-Penrose (NP) formalism. In particular, the effective potential for the Zerilli equation (ZE) is analyzed and the existence of the Dirac BQNMs is proved. The entropy and area spectra of the ZZLBH are shown to be evenly spaced.

The characteristic resonance spectra of the confined scalar fields in the RLDBH geometry are studied analytically. The KGE is solved and the boxed QNMFs (BQNMFs) of the caged RLDBH are obtained. Employing MM, the entropy and area spectra of the RLDBH are investigated, as well as the wave dynamics of a charged massive scalar field propagating in a maximally RLDBH (MRLDBH) geometry by solving the Klein-Gordon-Fock equation (KGFE). The existence of a discrete and infinite family of resonances describing non-decaying scalar configurations enclosing the MRLDBHs, which can indicate the possible existence of dark matter distributions around them is proved. Particularly, the effective heights of those clouds above the center of the MRLDBH are analytically computed.

# ÖZ

Bu tezde sırasıyla, Garfinkle-Horowitz-Strominger karadeliği (GHSBH), sıfır dinamik katsayısına sahip dört boyutlu Lifshitz karadeliği (ZZLBH) ve dönen lineer dilaton karadeliğinin (RLDBH'nin) entropi ve alan spektrumlarını araştırmak için adyabatik değişmezde görünen geçiş frekansını aşırı sönümlü kuazinormal mod frekanslarından (QNMF'lerden) ölçen Maggiore'nin metodu (MM) kullanılmıştır. GHSBH'nin kuantizasyonunu çalışmak için, sıradan kuazinormal modlar (QNM'ler) yerine, GHSBH geometrisindeki sınırlı skaler alanların karakteristik rezonans spektrumları olan kutulu QNM'ler (BQNM'ler) hesaplanmıştır. Bu amaçla, GHSBH'nin olay ufkunun yakınında yer alan kafesleyici bir aynaya sahip olduğu varsayılmıştır. Daha sonra, bu kafesli karadeliğin (BH'nin) kompleks rezonans frekanslarının olay ufku etrafındaki skaler pertürbasyonlar göz önüne alınarak nasıl hesaplandığı gösterilmiştir. Ek olarak, GHSBH'nin rezonans modları (RM'leri) ve RM frekansları (RMF'leri) yüksek azimutal kuantum sayısına sahip sınırlı skaler dalgalar kullanılarak hesaplanmıştır. Entropi ve alan kuantizasyonlarının GHSBH parametrelerinden bağımsız olduğu gösterilmiştir, ancak her iki spektrum da eşit aralıklıdır.

Klein-Gordon denklemi (KGE) çözilerek ZZLBH'nin spektroskopisi incelenmiştir. Adyabatik değişmez nicelik kullanılarak, masif skaler dalgaların olay ufkundaki Schrödinger benzeri denklemin (SLE) tam çözümüne dayanarak, ZZLBH'nin QNM'leri hesaplanmıştır. Bu uzay-zamanın fermiyonik pertürbasyonları için Dirac denklemleri, Newman-Penrose (NP) formalizmi çerçevesinde çözülmüştür. Özellikle, Zerilli denklemi (ZE) için etkili potansiyel analiz edilmiş ve Dirac BQNM'lerinin varlığı kanıtlanmıştır. ZZLBH'nin entropi ve alan spektrumlarının eşit aralıklı olduğu gösterilmiştir.

RLDBH geometrisindeki sınırlı skaler alanların karakteristik rezonans spektrumları analitik olarak incelenmiştir. Kafesli RLDBH'nin kutulu QNMF'leri (BQNMF'leri) elde edilmiştir. MM kullanılarak, RLDBH'nin entropi ve alan spektrumlarının yanı sıra, maksimal bir RLDBH (MRLDBH) geometrisinde ilerleyen yüklü bir masif skaler alanın dalga dinamiği Klein-Gordon-Fock denklemi (KGFE) çözülerek incelenmiştir. Etraflarında olası karanlık madde dağılımları bulunduğuna işaret eden, MRLDBH'leri çevreleyen azalmayan skaler konfigürasyonları tanımlayan ayrık ve sonsuz bir rezonans setinin varlığı gösterilmiştir. Özellikle, bu bulutların MRLDBH'nin merkezi üzerindeki efektif yükseklikleri analitik olarak hesaplanmıştır.

**Anahtar Kelimeler:** Garfinkle-Horowitz-Strominger kara deliği, Lifshitz kara deliği, dönen doğrusal dilaton karadeliği, karanlık madde, sınırlayıcı kafes, skaler ve fermiyonik alan pertürbasyonları, Klein-Gordon denklemi, Klein-Gordon-Fock denklemi, Dirac denklemi, Newman-Penrose formalizmi, skaler bulut, kuazinormal mod, kutulu kuazinormal mod, rezonans modu, kara delik spektroskopisi, entropi ve alan kuantizasyonu.

To All My Loved Ones

# ACKNOWLEDGMENT

I would like to express my deepest appreciation to Prof. Dr. İzzet Sakallı (Chairman of Physics and Chemistry Departments), who is my supervisor, for providing me continuous support in various ways, encouragement and guidance in the course of my postgraduate studies including the preparation of this thesis. Besides, I would also like to thank all the members of Physics and Chemistry Departments; the esteemed former chairman Prof. Dr. Mustafa Halilsoy, the instructors, and Mrs. Çilem Aydınlık and Mr. Reşat Akoğlu for their helpful advices and support during my graduate study.

I sincerely thank all the jury members, for their participation to my defense as well as their suggestions and comments for the improvement of my thesis.

I am also grateful to my loving friends; Sara Kanzi and Danial Forghani from Physics Department, Mamoon Alokour and Abdallah Alshhab from Chemistry Department, and my old friends Ceren Demirtaş and Muratcan Yüce; who have always been there to help and motivate me morally.

Above all, I would especially like to express how grateful I am to my precious parents Burçin Bayraktar Tokgöz and Serhat Tokgöz, and my lovely brothers Ali Rıza Tokgöz and Rıza Orhun Tokgöz, for their eternal love and everything that they have done to support me.

# TABLE OF CONTENTS

# LIST OF TABLES

# LIST OF FIGURES

# LIST OF ABBREVIATIONS

| | |
|---|---|
| AdS | Anti-de Sitter |
| AF | Asymptotically Flat |
| BH | Black Hole |
| BP | Barrier Peak |
| BQNM | Boxed Quasinormal Mode |
| BQNMF | Boxed Quasinormal Mode Frequency |
| BSQR | Bohr-Sommerfeld Quantization Rule |
| BY | Brown-York |
| CDE | Chandrasekhar-Dirac Equation |
| CG | Conformal Gravity |
| CH | Confluent Hypergeometric |
| DBC | Dirichlet Boundary Condition |
| dS | de Sitter |
| EMDA | Einstein-Maxwell-Dilaton-Axion |
| GHSBH | Garfinkle-Horowitz-Strominger |
| GR | General Relativity |
| KGE | Klein-Gordon Equation |
| KGFE | Klein-Gordon-Fock Equation |
| MM | Maggiore's Method |
| MRLDBH | Maximally Rotating Linear Dilaton Black Hole |
| NAF | Non-Asymptotically Flat |
| NBC | Neumann Boundary Condition |
| NH | Near-Horizon |
| NP | Newman-Penrose |
| QGT | Quantum Gravity Theory |
| QM | Quantum Mechanics |
| QNM | Quasinormal Mode |
| QNMF | Quasinormal Mode Frequency |
| RLDBH | Rotating Linear Dilaton Black Hole |
| RM | Resonance Mode |
| RMF | Resonance Mode Frequency |
| RN | Reissner-Nordström |
| SLE | Schrödinger-Like Equation |
| WKB | Wentzel-Kramers-Brillouin |
| ZE | Zerilli Equation |
| ZZLBH | Lifshitz Black Hole with Zero Dynamical Exponent |

# Chapter 1

# INTRODUCTION

Quantum mechanics (QM) and general relativity (GR) fail when confronted with each other's principles and are therefore limited in their ability to describe the Universe. Recent developments in physics show that our universe has a more complex structure than that predicted by the standard model [1]. Currently, one of the greatest projects in theoretical physics is to unify GR with QM to make them compatible with each other. The resulting new unified theory is expected to be an important tool that can tackle the problem in describing the behavior of the Universe, from leptons and quarks to galaxies. This is the quantum gravity theory (QGT) [2]. However, current QGT still requires further extensive development, it is not completed yet.

The effects of gravity are very strong near the BHs and quantum effects could not be ignored in the NH region. QGT [3] seeks to describe gravity according to the principles of QM. One of the difficulties of formulating the QGT is that quantum gravitational effects only appear at length scales near the Planck scale, around $10^{-35}$ meter, a scale far smaller, and equivalently far larger in energy, than those currently accessible by high energy particle accelerators. Therefore, we physicists lack experimental data.

The onset of the QGT dates back to the seventies when Hawking [4,5] and Bekenstein [6-10] amalgamated GR and QM by using the BH as a theoretical arena and showed that the BH is a quantum mechanical system rather than a classical object. Theoretical studies on BHs have attracted substantial attention since the advent of GR. It is well known that BHs cannot be directly observed but they can be detected through their action on their neighborhoods. The behavior of the matter and fields surrounding a BH not only tells us about its presence but also help us determine its parameters. A realistic BH can never be fully described by its basic parameters (mass, charge and angular momentum) and is always in the perturbed state [11].

The quantization of BHs was first proposed in the seminal works of Bekenstein [7,8] in which the quantization procedure is based on the surface area $\mathcal{A}$ (so the entropy $S^{BH}$) of a BH that acts as a classical adiabatic invariant. According to the Ehrenfest principle [12] (There is a proportionality between $S^{BH}$ and $\mathcal{A}$: $S^{BH} = \frac{\mathcal{A}}{4\hbar}$, which is attested from the adiabatic invariance [13].), any classical adiabatic invariant should also have a quantum entity with a discrete spectrum. In the spirit of Ehrenfest's adiabatic hypothesis [13,14], Bekenstein [7,8] hypothesized that the area of a quantum BH should only have the following discrete equidistant spectrum:

$$\mathcal{A}_n = \varepsilon n\hbar = 8\pi\xi n\hbar, \qquad n = 0, 1, 2, 3, \ldots \qquad (1)$$

where $\varepsilon$ is known as the unknown fudge factor (undetermined dimensionless constant) [15] and $\xi$ is the order of unity. One can immediately deduce from the above expression that the minimum increase in the horizon area should be $\Delta\mathcal{A}_{min}$. The best-known values of $\varepsilon$ are $8\pi$ (or $\xi = 1$) [8,9,16] for the Schwarzschild BH and also for the Kerr-Newman BH, and $4lnp$ where $p = 2,3,\ldots$ [17-20]. Namely, the BH horizon is formed by patches of equal area [8].

Based upon the seminal works of Bekenstein, many attempts have been made in order to study the quantum spectrum of the BHs. As a result, different spectra have been obtained. Some methods used for obtaining the spectrum can admit the value of the coefficient $\varepsilon$ different than that obtained by Bekenstein; this has led to the discussion of this subject in the literature (for a review of this topic, refer to [21] and references therein). Maggiore's result [16] is the one which shows a perfect agreement with Bekenstein's result by modifying the formula of Kunstatter [22] as follows:

$$I_{adb} = \int \frac{dM}{\Delta \omega}, \tag{2}$$

where $I_{adb}$ denotes the adiabatic invariant quantity and $\Delta \omega = \omega_{n+1} - \omega_n$ represents the transition frequency between the subsequent levels of an uncharged and static BH having the total energy (or mass) $M$. However, the researchers [23-25] working on this issue later realized that the above definition is not suitable for the hairy BHs (massive, charged, and rotating ones) that the generalized form of the definition should be given by (see [26] and references therein):

$$I_{adb} = \int \frac{T_H dS^{BH}}{\Delta \omega}, \tag{3}$$

where $T_H$ is the temperature and $S^{BH}$ denotes the entropy of the BH. Thus, using the first law of BH thermodynamics, the above equation can be modified for the considered BH. On the other hand, the Bohr-Sommerfeld quantization rule (BSQR) [27] states that $I_{adb}$ acts as a quantized quantity ($I_{adb} \simeq n\hbar$) while the quantum number $n$ tends to infinity.

For several decades, many efforts are made in investigating the evolution of external field perturbations around the BHs which lead to damped oscillations called QNMs [28-30]. Different kinds of perturbations in a BH geometry can excite certain combination of its characteristic frequencies of the normal modes for which the frequencies are no longer purely real showing that the system is losing energy. The real part of QNMF represents the ring down frequency and the imaginary part, the decay time. The resonance frequencies of the response of external perturbations (called QNMFs) are one of the most essential characteristics of a BH. It is widely believed that the QNMs carry a unique characteristic fingerprint which would lead to the direct identification of the BH existence. QNMs associated with perturbations of different fields were considered in a variety of works [31-36].

In 2008, Maggiore [16] postulated that a BH can be viewed as a damped harmonic oscillator whose physically relevant frequency is identical to the complex QNMFs having both real and imaginary parts. To obtain $\Delta\omega$, Maggiore [16] used the fact that for the perturbed BH's complex QNMF which is in the form of $\omega = (\omega_R^2 + \omega_I^2)^{\frac{1}{2}}$ in which $\omega_R$ and $\omega_I$ are the real and imaginary parts of the frequency, respectively; the imaginary part is dominant over the real part ($\omega_I \gg \omega_R$) for the ultrahigh dampings ($n \to \infty$) implying that $\Delta\omega \simeq \Delta\omega_I = |\omega_I|_{n+1} - |\omega_I|_n$. He activated Kunstatter's formula and used the BSQR [27] to prove that for the Schwarzschild BH the area spectrum is exactly equal to Bekenstein's original result [8]. Meanwhile, one of the most important contributions was made by Hod [19,37], who was the first physicist to argue that the QNMs [38,39] or the so-called ringing modes (characteristic sound of BHs and neutron stars [40]) for the highly excited modes can be used in the identification of the quantum transitions for the $I_{adb}$. He suggested that $\varepsilon$ could be determined by utilizing the QNMFs [41] such that the real part of the asymptotic QNMFs are considered as a transition frequency (the smallest energy a BH can emit) in the semiclassical limit, and in sequel its quantum emission gives rise to a change in the BH mass, which is related to $\mathcal{A}_n$. In this way, he found that $\varepsilon = 4 \ln 3$ (for a Schwarzschild BH) [19]. Subsequently, MM has been employed in numerous studies of BH quantization in the literature and Bekenstein's conjecture has been tested for various BH solutions (see, for instance, [16,23,24,42-57]). Today, there are numerous methods in the literature to compute the QNMs, such as the WKB method, the phase integral method, continued fractions and direct integrations of the wave equation in the frequency domain [58].

According to the no-hair conjecture [19,59,60] which is one of the milestones in understanding the subject of BH physics [61,62], BHs are the fundamental objects like the atoms in QM and they should be characterized by only 3 -parameters: mass, charge, and angular momentum. In fact, the no-hair conjecture has an oversimplified physical picture: all residual matter fields for a newly born BH would either be absorbed by the BH or be radiated away to spatial infinity (this scenario excludes the fields having conserved charges) [19,63,64]. In accordance with the same line of thought, other no-hair theorems indeed omitted static spin-0 fields [65-68], spin-1 fields [69-72], and spin-1/2 fields [73,74] from the exterior of stationary BHs. On the other hand, the significant developments in theoretical physics have led to the other types of hairy BH solutions. A notable example is the colored BHs in [75,76]. In addition to the mass, for full characterization, a colored BH needs an additional integer number (independent of any conserved charge) which is assigned to the nodes of the Yang-Mills field. Other hairy BHs include different types of exterior fields that belong to the Einstein-Yang-Mills-Dilaton, Einstein-Yang-Mills-Higgs, Einstein-Skyrme, Einstein-Non-Abelian-Proca, Einstein-Gauss-Bonnet etc. theories (see, for example, [77-96]). Interestingly, many no-hair theories [7,63,68,71-74,97,98] do not cover the time-dependent field configurations surrounding the BH. Astronomical BHs are not tiny and unstable, but very heavy, large and practically indestructible. Observations show that in the densely populated center of most galaxies, including ours, there are monstrous BHs [99], which are many hundreds of millions of times heavier than the Sun. As was shown in [100], the regular time-decaying scalar field configurations surrounding a supermassive Schwarzschild BH do not fade away in a short time (according to the dynamical chronograph governed by the BH mass). Besides, ultra-light scalar fields are considered as a possible candidate for the dark matter halo (see, for instance, [100-103]).

Hod [64] extended the outcomes of [100] to the Kerr BH, which is well-suit for studying astrophysical wave dynamics. Hod proved the presence of an infinite family of resonances (discrete) describing non-decaying (stationary) scalar configurations surrounding maximally rotating Kerr BH. Thus, contrary to the finite lifetime of the static regular scalar configurations mentioned in [100], the stationary and regular scalar field configurations (clouds) surrounding the realistic rotating (Kerr) BH survive infinitely long [64]. To this end, Hod considered the dynamics of massive KGE in the Kerr geometry. Moreover, the effective heights of those clouds above the center of the BH were analytically computed. The obtained results support the lower bound conjecture of Núñiez et al. [98].

In this thesis, we mainly focus on the investigation of the entropy/area quantization (spectroscopy) of three BHs: the GHSBH, the ZZLBH, and the RLDBH, respectively; and the computation of the stationary resonances together with the effective heights of the scalar clouds of the MRLDBH.

In Chapter 2, we focus on the investigation of the entropy/area quantization (spectroscopy) of the GHSBH [104]. GHSBH is a member of a family of solutions to the low-energy limit of the string theory whose action is obtained when the Einstein-Maxwell theory is enlarged to involve a dilaton field $\phi$, which couples to the metric and the gauge field non-trivially. That is why the physical properties of the charged stringy BHs differ significantly from the Reissner-Nordström (RN) BH. To employ the MM, the QNMs (a set of complex frequencies arising from the perturbed BH) of the GHSBH should be computed. To achieve this, we first consider the KGE for a massless scalar field in the background of the GHSBH. After separating the angular and the radial equations, we obtain a SLE which is the so-called ZE [61]. In fact, the spectroscopy problem of the GHSBH was first studied by Wei et al. [25]. In the adiabatic invariant quantity, they used the ordinary QNMs of Chen and Jing [105] who computed the associated QNMs to obtain an equal spacing of GHSBH spectra at the high frequency modes by using the monodromy method [58].

Our first goal in Chapter 2 [106] is to reconsider the problem of GHSBH spectroscopy by using a recent analytical method, which is invented by Hod [107] for obtaining the GHSBH's resonance spectra and computing the BQNMs [108-111], instead of the ordinary QNMs, and is different from the monodromy method. Thus, we seek to support the study of Wei et al. [25] because we believe that the studies that obtain the same conclusion using different methods are more reliable. Hod's idea is indeed based on the recent study [112], which provides compelling evidence that the confined scalar fields in a cavity of the coupled Einstein-Klein-Gordon system collapse to form caged BHs, generically. To this end, we assume that there exists a mirror or a finite-volume cavity confining the GHSBH that is placed at a constant radial coordinate with a radius $r_m$, which is in the vicinity of the horizon. The scalar field $\Phi$ is imposed to terminate at the location of the mirror, which requires to use both the Dirichlet boundary condition (DBC) and Neumann boundary condition (NBC). In that scenario, the radial wave equation was studied around the NH region. In the framework of this scenario, we focus our analysis of the radial wave equation on the NH region [107]. We then derive the BQNMs of the GHSBH based on the fact that, for the QNMs to exist, the outgoing waves must be terminated at the event horizon. The NH form of the ZE is reduced to a Bessel differential equation [113]. After choosing the expedient solution, we impose the DBC and NBC. Then, we consider some of the transformed features of the Bessel functions for finding the resonance conditions. Next, we use an iteration scheme to define the BQNMs of the GHSBH. Once the BQNMs are computed, we use the transition frequency in the adiabatic invariant and obtain the GHSBH area/entropy spectra.

As a second goal in this chapter, we want to reconsider the same problem with another scenario. The effective potential generated from the massless KGE performs a barrier peak (BP) for the propagating scalar waves when the azimuthal quantum number $l$ gets high values. As a result, the scalar waves are self-confined between the horizon and the BP. This will yield characteristic RMs of the confined scalar fields in the GHSBH geometry. To this end, the scalar field is imposed to be terminated at the BP and to be purely ingoing wave at the horizon. In fact, our method is similar to those of the researches [45,48,51,114] which are mainly inspired from the studies [115,116] in which the QNMs are computed using the poles of the scattering amplitude in the Born approximation. After reducing the radial KGE to the one-dimensional SLE [61], we show that it becomes a confluent hypergeometric (CH) differential equation [113] in the NH region. Imposing the relevant boundary condition and then using the pole feature of the Gamma function which appears in the solution, we obtain the RMFs of the GHSBH. By using the highly damping RMs in the MM, we get the entropy/area spectra of the GHSBH.

Chapter 3 mainly explores the entropy/area spectra of a ZZLBH [117] possessing a particular dynamical exponent $z=0$. In fact, this problem was previously studied [118,119]. The QNMs were calculated and it was shown that the Lifshitz BH possesses a discrete and equidistant spectrum under scalar field perturbations. The Lifshitz spacetimes have attracted attention from the researchers working on condensed matter and quantum field theories [120] for being invariant under anisotropic scale and characterizing gravitational dual of strange metals [121].

Our first aim in this chapter is to perform the QNM calculations of the ZZLBH in order to implement MM successfully. To this end, we consider the KGE for a massive scalar field in the ZZLBH background to find the QNMs. Separation of the angular and the radial equations yields a SLE [61]. Asymptotic limits of the potential (38) show that the effective potential may diverge beyond the BH horizon for the massive scalar particle; thus, in the far region, the QNMs might not be perceived by the observer. Hence, following the particular method prescribed in [45,51,114-116], we focus our analysis on the NH region and impose the boundary conditions (121) that outgoing waves to vanish at the event horizon and waves to be null at spatial infinity. After getting NH form of the SLE, we show that the radial equation is reduced to a CH differential equation [113]. Performing some manipulations on the NH solution and using the pole structure of the Gamma functions [113], we show how one finds out the QNMs as in [45,49,114,122-124]. The imaginary part of the QNMs is used in $I_{adb}$, and the quantum spectra of entropy and area of the ZZLBH are obtained.

The interaction of the BHs (depending on the couplings of spin-rotation, field mass-BH mass, field charge- BH charge) with a fermionic (Dirac) field has attracted deep theoretical interest and it has motivated the researchers to have a better understanding of various physical fields, especially the spin-1/2 particles, in the vicinity of the BHs. As a second task in Chapter 3, we reconsider the problem with fermionic perturbations and instead of the ordinary QNMs, we derive the BQNMs [52,107,109-111] that are the characteristic resonance spectra of the confined scalar fields in the ZZLBH geometry [117]. We solve the Dirac equation with $q = 0$ and $\mu^* = 0$ for spin-1/2 test particles in the ZZLBH background. The Dirac equations are decoupled into a radial set and an angular set within NP formalism. The separation constant is obtained with the aid of the spin weighted spheroidal harmonics. The radial set of equations, which is independent of mass, is reduced to a set of SLEs or the so-called ZEs [61] with their NH forms. To this end, we consider a mirror (confining cavity) surrounding the ZZLBH which is located at a constant radial coordinate. Next, we impose that the fermionic field should terminate at the mirror's location, which requires two boundary conditions to be used: DBC and NBC [52,106,107,112]. With this scenario, we focus our analysis of the ZE in the NH region [107]. In the NH region, these equations with their associated Zerilli potentials are solved in terms of the Bessel functions of the first and second kinds [113] arising from the fermionic perturbation on the background geometry. For computing the BQNMs instead of the ordinary QNMs, we first impose the purely ingoing wave condition at the event horizon. And then, DBC and NBC are applied in order to get the resonance conditions. For solving the resonance conditions, we follow an iteration method. We show the existence of the Dirac BQNMs of ZZLBH using the iteration scheme in Chapter 2. Finally, MM is employed to derive the entropy/area spectra of the ZZLBH which are shown to be equidistant.

In Chapter 4, we introduce the RLDBH and study its spectroscopy [125]. These BHs have a NAF structure, similar to our universe model: Friedmann-Lemaître-Robertson-Walker spacetime [126]. RLDBHs are the solutions to the Einstein-Maxwell-dilaton-axion (EMDA) theory [127].

Primarily, in order to do the quantization analysis on the RLDBH background, we use MM [16], however, instead of the ordinary QNMs (or vibrational modes) we consider the BQNMs (also known as quasi-bound states) [107] which reveal when the considered BH is caged [112]. Namely, RLDBH is assumed to be confined in a finite-volume cavity that can be achieved by placing an artificial spherical reflecting surface (mirror) at some distance $r_m$. This configuration is akin to the perfect mirror (used by Press and Teukolsky [128]) to perform a BH bomb. However, for studying the resonance frequencies of the caged BHs, one can follow the recent studies [106,107] in which $r_m$ was chosen to be close to the NH region [129]. The propagating scalar fields $\Psi$ are imposed to vanish at $r_m$. This requirement is fulfilled by using two boundary conditions: the DBC ($\Psi(r)|_{r=r_m} = 0$) and the NBC ($\frac{d\Psi(r)}{dr}\big|_{r=r_m} = 0$). In addition to this, one should use the fact that QNMs are the pure ingoing waves at the event horizon. These three conditions guide us to find the resonance condition, which yields the BQNMs after some manipulations. Here, we will use the transition frequency of the BQNMs in Eq. (2) to derive the RLDBH's spectroscopy (area/entropy spectra). It is worth mentioning that the spectroscopy of RLDBH has recently been studied by Sakalli [50] via the standard QNMs, and the present study gives support to that paper.

Secondly, in the line with the study by [64], our next purpose in Chapter 4 is to explore the stationary and regular scalar field configurations surrounding a MRLDBH [125]. These BHs include dilaton and axion fields, which are candidates for the dark matter halo. Some current experiments are focused on the relationship between the dark matter and the dilaton and axion fields [130-133]. Various studies have also focused on the RLDBHs [50,52,134-136]. We consider the charged massive KGFE [137-139] in the MRLDBH geometry. Thus, we investigate wave dynamics in that geometry and seek for the existence of possible resonances describing the stationary charged and massive scalar field configurations surrounding the MRLDBH.

The thesis is arranged as stated in the following elaborations. In Chapter 2, we introduce the GHSBH metric and study the KGE for a massless scalar field in this geometry. Then, we reduce the physical problem to the ZE by using the separation of variables technique. Firstly, we show that the ZE reduces to a Bessel differential equation in the vicinity of the event horizon. The DBC and NBC at the surface $r = r_m$ of the confining cavity single out two discrete sets of complex BQNMs of the caged GHSBH. We apply MM to obtain the quantum spectra of the entropy/area of the GHSBH. Secondly, we reconsider the same problem with another scenario and we show how the SLE reduces to a CH differential equation. After computing of the RMs of the GHSBH, we apply MM to obtain the GHSBH's spectroscopy.

In Chapter 3, we briefly review the ZZLBH geometry and its physical properties. First, we show the separation of the massive KGE and find the effective potential. Next, we solve the NH SLE and show how QNMs are calculated. Then, we compute the entropy/area spectra of the ZZLBH. Second, in the framework of NP formalism, we solve the Dirac equations with uncharged massless fermionic test particles in the ZZLBH spacetime. Particularly, we analyze the effective potential for the ZE and show the existence of the Dirac BQNMs of the ZZLBH.

In Chapter 4, we overview the RLDBH metric briefly with its characteristic properties as well as the MRLDBH metric. Firstly, we analyze the KGE for a massless scalar field, and derive the ZE in this geometry. We show the reduction of the ZE to a Bessel differential equation around the NH. Next, we impose the required boundary conditions for computing the complex BQNMs of the caged RLDBH. Then, applying MM, we obtain the spectroscopy of RLDBH. And secondly, we study the charged massive scalar field perturbation and separate the KGFE in the MRLDBH geometry. We explore the existence of a discrete family of resonances describing stationary scalar configurations surrounding MRLDBH, and in sequel we compute the effective heights of those scalar configurations above the central MRLDBH.

Finally, we present our conclusions and future scopes in Chapter 5. Throughout the thesis, the geometrized unit system, in which the fundamental constants are given as $G = c = k_B = 1$ and $\ell_\rho^2 = \hbar$, is used. However, in section 3.1, we set the natural units with $G = c = k_B = \hbar = 1$.

# Chapter 2

# SPECTROSCOPY OF THE GHSBH[1]

## 2.1 GHSBH Spacetime

In this section, we represent the geometry and some of the thermodynamical properties of the GHSBH [104]. In the low-energy limit of the string field theory, the four-dimensional Einstein-Maxwell-dilaton low-energy action [15], which defines the coupling between the dilaton field $\phi$ and $U(1)$ gauge field is expressed in Einstein frame as

$$S = \int d^4x \sqrt{-g}\left[-R + 2(\nabla\phi)^2 + e^{-2\phi}F^2\right], \qquad (4)$$

with $F^2 = F_{\mu\nu}F^{\mu\nu}$ in which $F_{\mu\nu}$ is the Maxwell field associated with a $U(1)$ subgroup of $E_8 \times E_8$ (or Spin (32)/$Z_2$) [104]. The field equations are obtained, by applying the variational principle to the action (4), as follows:

$$\nabla_\mu(e^{-2\phi}F^{\mu\nu}) = 0, \qquad (5)$$

$$\nabla^2\phi + \frac{1}{2}e^{-2\phi}F^2 = 0, \qquad (6)$$

$$R_{\mu\nu} = 2\nabla_\mu\phi\nabla_\nu\phi - g_{\mu\nu}(\nabla\phi)^2 + 2e^{-2\phi}F_{\mu\rho}F_\nu^{\ \rho} - \frac{1}{2}g_{\mu\nu}e^{-2\phi}F^2. \qquad (7)$$

The GHSBH is the solution to the action (4) and it has the following static and spherically symmetric metric:

$$ds^2 = -f(r)dt^2 + \frac{dr^2}{f(r)} + g(r)d\Omega^2, \qquad (8)$$

---

[1] This chapter is mainly quoted from [106,149].

where $d\Omega^2$ is the standard metric on the 2-sphere. The metric functions in (8) are given by

$$f(r) = \frac{r - r_+}{r}, \tag{9}$$

$$g(r) = r^2 - 2ar. \tag{10}$$

The physical parameter $a$ is defined by

$$a = \frac{Q^2}{2M} e^{-2\phi_0}, \tag{11}$$

where $M$, $Q$ and $\phi_0$ describe the mass, magnetic charge and the asymptotic value (constant) of the dilaton, respectively; whereas $r_+ = 2M$ corresponds to the event horizon of the GHSBH. Note that the mass calculation of the GHSBH was done in [140] with Komar's mass integral formulation [141]. It is worth noting that in this spacetime, the dilaton field reads

$$e^{-2\phi} = \left(1 - \frac{2a}{r}\right) e^{-2\phi_0}, \tag{12}$$

and the Maxwell field is given by

$$F = Q \sin\theta \, d\theta \wedge d\varphi. \tag{13}$$

On the other hand, it is straightforward to get the electric charge case by applying the following duality transformations:

$$\tilde{F}_{\mu\nu} = \frac{1}{2} e^{-2\phi} \varepsilon_{\mu\nu}^{\alpha\beta} F_{\alpha\beta}, \quad \phi \to -\phi. \tag{14}$$

Because the GHSBH and the Schwarzschild BH have the same $R^2$ part, their surface gravities [142] coincide with each other:

$$\kappa = \lim_{r \to r_+} \sqrt{-\frac{1}{2} \nabla_\mu \chi_\nu \nabla^\mu \chi^\nu} = \frac{1}{2} \frac{df(r)}{dr}\bigg|_{r=r_+} = \frac{1}{4M}, \tag{15}$$

where $\chi^\nu$ is the timelike Killing vector defined as $\chi^\nu = [1,0,0,0]$. Therefore, one can easily obtain the Hawking temperature $T_H$ of the GHSBH as

$$T_H = \frac{\hbar \kappa}{2\pi} = \frac{\hbar}{8\pi M}. \tag{16}$$

Thus, $T_H$ of the GHSBH does not depend on the charge. The areal radii of the GHSBH and the Schwarzschild BH are different, accordingly their areas and entropies. So, the entropy of the GHSBH reads

$$S^{BH} = \frac{\mathcal{A}}{4\hbar} = \frac{\pi r_+(r_+ - 2a)}{\hbar}. \tag{17}$$

Actually, when $a = M$ (in case of extremal charge: $Q = \sqrt{2}Me^{\phi_0}$), the GHSBH's area vanishes, so its entropy, too. It is nothing but a naked singularity. Moreover, for having physical (nondecreasing) entropy, it is necessary to have $a \leq M$. The singularity of GHSBH is null (unlike to the timelike singularity of RN BH) and therefore outward radial null geodesics do not hit it (see, for instance, [39]). On the other hand, the first law of thermodynamics for the GHSBH takes the following form:

$$T_H dS^{BH} = dM - U_H dQ, \tag{18}$$

in which $U_H = aQ^{-1}$ is the electric potential on the horizon.

## 2.2 Quantization of the Caged GHSBH

In this section, we separate the KGE for a massless scalar field propagating in the GHSBH background and derive the ZE with its associated effective potential in the NH region. We compute the BQNMFs to obtain the entropy/area spectra of the caged GHSBH [106].

### 2.2.1 Separation of the Massless KGE on the Caged GHSBH

We may start the spectroscopy analysis of the GHSBH from the BQNMs via MM, regarding the massless scalar field $\Psi$ which fulfills the KGE:

$$\frac{1}{\sqrt{-g}} \partial_\mu (\sqrt{-g} g^{\mu\nu} \partial_\nu \Psi) = 0. \tag{19}$$

We shall choose the ansatz for the scalar field $\Psi$ in the form:

$$\Psi = g(r)^{-1/2} H(r) e^{-i\omega t} Y_l^m(\theta, \phi), \quad Re(\omega) > 0, \tag{20}$$

where $\omega$ is the frequency of the propagating scalar wave and $Y_l^m(\theta, \phi)$ represents the spheroidal harmonics with the eigenvalue $L = -l(l+1)$. Here, $l$ and $m$ are the azimuthal quantum number and the magnetic quantum number, respectively.

After some straightforward algebra, one can reduce the radial equation to the one-dimensional SLE or the so-called ZE [61] in the following form:

$$\left[-\frac{d^2}{dr^{*2}} + V(r)\right] H(r) = \omega^2 H(r), \tag{21}$$

where $r^*$ represents the tortoise coordinate which is defined by

$$r^* = \int \frac{dr}{f(r)}. \tag{22}$$

Evaluating the above integral, we get

$$r^* = r + r_+ \ln\left(\frac{r - r_+}{r_+}\right). \tag{23}$$

Furthermore, one can inversely obtain

$$r = r_+[1 + W(u)], \tag{24}$$

where $u = e^{\left(\frac{r^* - r_+}{r_+}\right)}$, and $W(u)$ is the Lambert-W or the so-called omega function [38]. The tortoise coordinate has the following limits:

$$\lim_{r \to r_+} r^* = -\infty, \tag{25}$$

$$\lim_{r \to \infty} r^* = \infty. \tag{26}$$

The Zerilli potential $V(r)$ seen in the ZE is found to be

$$V(r) = \frac{f(r)}{r(r - 2a)}\left[L - \frac{a^2}{r(r - 2a)} f(r) + \frac{r_+(r - a)}{r^2}\right]. \tag{27}$$

### 2.2.2 BQNMFs and Entropy/Area Spectra of the Caged GHSBH

Our interest in this section is to solve the ZE (21) around the NH. We impose the QNM boundary condition that only ingoing plane waves are allowed at the event horizon, in order for computing the BQNMFs. Then, we apply the DBC and NDC for having the resonance conditions, borrowing ideas from Hod's recent study [107]. Finally, in computing the BQNMs, we use an iteration scheme to solve the resonance conditions. To do so, we rewrite the metric function (9) as follows

$$f(r) \to f(x) = \frac{x}{x+1}, \tag{28}$$

where

$$x = \frac{r-r_+}{r_+}. \tag{29}$$

Thus, one can find the series expansion of $f(x)$ in the leading order as follows

$$f(x) \cong x + O(x^2), \tag{30}$$

and express $r^*$ in the NH region ($x \to 0$) as

$$r^* = \int \frac{r_+ dx}{f(x)} \cong r_+ \ln(x) = \frac{1}{2\kappa} \ln(x). \tag{31}$$

From Eq. (31), one also reads

$$x = e^{2y}, \tag{32}$$

with

$$y = \kappa r^*. \tag{33}$$

Substituting Eq. (29) into Eq. (27), the NH Zerilli potential is approximated by

$$V_{NH}(x) = \frac{L+\beta}{r_+ \gamma} x + O(x^2), \tag{34}$$

with the parameters given below:

$$\beta = \frac{r_+ - a}{r_+}, \tag{35}$$

$$\gamma = r_+ - 2a. \tag{36}$$

After, substituting Eqs. (32)-(34) into Eq. (21), NH ZE can be obtained as

$$\left[ -\frac{d^2}{dy^2} + \frac{4r_+(L+\beta)}{\gamma} e^{2y} \right] H(y) = \widetilde{\omega}^2 H(y), \tag{37}$$

which has two linearly independent solutions given by

$$H(y) = C_1 J_{-i\widetilde{\omega}}\left(2i\sqrt{\Delta}e^y\right) + C_2 Y_{-i\widetilde{\omega}}\left(2i\sqrt{\Delta}e^y\right), \tag{38}$$

and identically

$$H(x) = C_1 J_{-i\tilde{\omega}}(2i\sqrt{\Delta x}) + C_2 Y_{-i\tilde{\omega}}(2i\sqrt{\Delta x}), \tag{39}$$

where $J_\nu(z)$ and $Y_\nu(z)$ are the Bessel functions [113] of the first and second kinds, with constants $C_1$ and $C_2$, respectively. The parameters of the Bessel functions read

$$\tilde{\omega} = \frac{\omega}{\kappa}, \tag{40}$$

$$\Delta = \frac{r_+(L+\beta)}{\gamma}. \tag{41}$$

For our analysis, we need the following limiting forms (when $\nu$ is fixed and $z \to 0$) of the Bessel functions [113,143]:

$$J_\nu(z) \sim \frac{[(1/2)z]^\nu}{\Gamma(1+\nu)}, \qquad \nu \neq -1, -2, -3, \ldots, \tag{42}$$

$$Y_\nu(z) \sim -\frac{1}{\pi}\Gamma(\nu)[(1/2)z]^{-\nu}, \qquad \Re\nu > 0. \tag{43}$$

Applying the above limiting forms, one can obtain the NH behavior ($e^y \ll 1$) of the solution (38) as

$$H \sim C_1 \frac{(i\sqrt{\Delta})^{-i\tilde{\omega}}}{\Gamma(1-i\tilde{\omega})} e^{-i\tilde{\omega} y} - C_2 \frac{1}{\pi}\Gamma(-i\tilde{\omega})(i\sqrt{\Delta})^{i\tilde{\omega}} e^{i\tilde{\omega} y}$$

$$= C_1 \frac{(i\sqrt{\Delta})^{-i\tilde{\omega}}}{\Gamma(1-i\tilde{\omega})} e^{-i\omega r^*} - C_2 \frac{1}{\pi}\Gamma(-i\tilde{\omega})(i\sqrt{\Delta})^{i\tilde{\omega}} e^{i\omega r^*}, \tag{44}$$

in which the constants $C_1$ and $C_2$ correspond to the amplitudes of the NH ingoing and outgoing waves, respectively. Since the QNM boundary condition implies that the outgoing waves must automatically vanish at the horizon, we conclude that the amplitude of the NH outgoing waves must be chosen as $C_2 = 0$. Thus, we are left with the acceptable solution of Eq. (39):

$$H(x) = C_1 J_{-i\tilde{\omega}}(2i\sqrt{\Delta x}). \tag{45}$$

Now, we consider the DBC at the surface $x = x_m$ of the confining cage [107,112]:

$$H(x)|_{x=x_m} = 0, \tag{46}$$

which in turn yields

$$J_{-i\tilde{\omega}}(2i\sqrt{\Delta x_m}) = 0. \qquad (47)$$

By using the following relation [113]

$$Y_v(z) = J_v(z)\cot(v\pi) - J_{-v}(z)\csc(v\pi), \qquad (48)$$

the condition (47) can be expressed as

$$\tan(i\tilde{\omega}\pi) = \frac{J_{i\tilde{\omega}}(2i\sqrt{\Delta x_m})}{Y_{i\tilde{\omega}}(2i\sqrt{\Delta x_m})}. \qquad (49)$$

The above equation is called the resonance condition. The boundary of the confining cage is located at the vicinity of the event horizon [107] according to the caged BH definition. Namely, in the NH region, we have

$$z_m \equiv \Delta x_m \ll 1 \quad \rightarrow \quad r_m \approx r_+. \qquad (50)$$

With the aid of Eqs. (42) and (43), the resonance condition (49) can be rewritten as follows:

$$\tan(i\tilde{\omega}\pi) \sim -\frac{\pi(i\sqrt{z_m})^{2i\tilde{\omega}}}{\Gamma(i\tilde{\omega})\Gamma(i\tilde{\omega}+1)} = i\frac{\pi e^{-\pi\tilde{\omega}}}{\tilde{\omega}\Gamma^2(i\tilde{\omega})}z_m^{i\tilde{\omega}}. \qquad (51)$$

On the other hand, the NBC is defined [107,112] by

$$\left.\frac{dH(x)}{dx}\right|_{x=x_m} = 0. \qquad (52)$$

Using the recurrence relations of the Bessel functions [113], we get

$$J_{-i\tilde{\omega}-1}(2i\sqrt{z_m}) - J_{-i\tilde{\omega}+1}(2i\sqrt{z_m}) = 0. \qquad (53)$$

Together with Eq. (48), one can derive the following relation:

$$Y_{v+1}(z) - Y_{v-1}(z) = \cot(v\pi)[J_{v+1}(z) - J_{v-1}(z)]$$
$$- \csc(v\pi)[J_{-v-1}(z) - J_{-v+1}(z)]. \qquad (54)$$

Furthermore, combining Eqs. (53) and (54), we find the resonance condition for NBC:

$$\tan(i\tilde{\omega}\pi) = \frac{J_{i\tilde{\omega}-1}(2i\sqrt{z_m})}{Y_{i\tilde{\omega}+1}(2i\sqrt{z_m})}\left[\frac{-1+J_{i\tilde{\omega}+1}(2i\sqrt{z_m})/J_{i\tilde{\omega}-1}(2i\sqrt{z_m})}{1-Y_{i\tilde{\omega}-1}(2i\sqrt{z_m})/Y_{i\tilde{\omega}+1}(2i\sqrt{z_m})}\right]. \qquad (55)$$

From the limiting forms given in Eqs. (42) and (43), in the NH region, we have

$$\frac{J_{i\widetilde{\omega}+1}(2i\sqrt{z_m})}{J_{i\widetilde{\omega}-1}(2i\sqrt{z_m})} \equiv \frac{Y_{i\widetilde{\omega}-1}(2i\sqrt{z_m})}{Y_{i\widetilde{\omega}+1}(2i\sqrt{z_m})} \sim O(z_m). \tag{56}$$

In this way, the resonance condition (55) takes the following form:

$$tan(i\widetilde{\omega}\pi) \sim \frac{J_{i\widetilde{\omega}-1}(2i\sqrt{z_m})}{Y_{i\widetilde{\omega}+1}(2i\sqrt{z_m})} = -i\frac{\pi e^{-\pi\widetilde{\omega}}}{\widetilde{\omega}\Gamma^2(i\widetilde{\omega})} z_m^{i\widetilde{\omega}}. \tag{57}$$

Since the resonance conditions (55) and (57) are small quantities, it is beneficial to use an iteration scheme in order to resolve the resonance conditions. Therefore, the $0^{th}$ order resonance equation is defined by [107]

$$tan\left(i\widetilde{\omega}_n^{(0)}\pi\right) = 0, \tag{58}$$

which means that

$$\widetilde{\omega}_n^{(0)} = -in, \quad n = 0, 1, 2, \dots. \tag{59}$$

After substituting Eq. (59) into r.h.s of Eqs. (55) and (57), the $1^{st}$ order resonance condition is obtained:

$$tan\left(i\widetilde{\omega}_n^{(1)}\pi\right) = \pm i\frac{\pi e^{i\pi n}}{(-in)\Gamma^2(n)} z_m^n. \tag{60}$$

which can be reduced to

$$tan\left(i\widetilde{\omega}_n^{(1)}\pi\right) = \mp n\frac{\pi(-z_m)^n}{(n!)^2}, \tag{61}$$

Here, minus (plus) stands for the DBC (NBC). To obtain the general characteristic resonance spectra of the caged GHSBH, we use the fact that in the $x \ll$ regime,

$$tan(x + n\pi) = tan(x) \approx x. \tag{62}$$

We have

$$i\widetilde{\omega}_n \pi = n\pi \mp n\frac{\pi(-z_m)^n}{(n!)^2}, \tag{63}$$

which gives

$$\widetilde{\omega}_n = -in\left[1 \mp n\frac{(-z_m)^n}{(n!)^2}\right]. \tag{64}$$

From the definition of the parameter $\widetilde{\omega}_n$ given in Eq. (40), the BQNMs read

$$\omega_n = -i\kappa n \left[1 \mp \frac{(-z_m)^n}{(n!)^2}\right], \qquad n = 0, 1, 2, ..., \tag{65}$$

where, $n$ is the resonance parameter or the so-called overtone quantum number [144]. The behavior of Eq. (65) for the highly excited states gives the following result:

$$\omega_n \approx -i\kappa n, \qquad n \to \infty. \tag{66}$$

which is in accordance with the results of [107,145-148]. Therefore, the transition frequency becomes

$$\Delta\omega_I = \kappa = \frac{2\pi T_H}{\hbar}. \tag{67}$$

Substituting the transition frequency (67) into Eq. (3), we have

$$I_{adb} = \frac{\hbar}{2\pi} S^{BH}. \tag{68}$$

Acting upon the BSQR ($I_{adb} = \hbar n$) [27], the entropy spectrum can be found as

$$S_n^{BH} = 2\pi n. \tag{69}$$

Since $S^{BH} = \mathcal{A}/4\hbar$, we can further read the area spectrum:

$$\mathcal{A}_n = 8\pi\hbar n. \tag{70}$$

Thence, the minimum area spacing reads

$$\Delta\mathcal{A}_{min} = 8\pi\hbar, \tag{71}$$

representing that the entropy/area spectra of the caged GHSBH are equispaced. Obviously, the spectra of the GHSBH are independent of the dilaton parameter $a$, and the spectral-spacing coefficient is $\varepsilon = 8\pi$, as the Bekenstein's original result [8-10]. Our result (71) also supports the study of Wei et al. [25].

## 2.3 Quantization of the GHSBH with Self-Confined Scalar Waves

In this section, different than the previous sections, we study the spectroscopy of the GHSBH from RMs instead of BQNMs [149]. Separation of the KGE (19) with the ansatz $\Psi = g(r)^{-\frac{1}{2}} H(r) e^{i\omega t} Y_l^m(\theta, \phi)$, $Re(\omega) > 0$ yields the ZE (21). As seen in the following sections, the effective potential (27) has a BP for high values of $l$. Therefore, we follow another particular method [45-47,53] and read the RMFs of the GHSBH.

**2.3.1 Potential Behavior of the GHSBH with Self-Confined Scalar Waves**

The azimuthal quantum number $l$ that appears in the effective potential given in Eq. (27), is a type of quantum number defined for an orbital which determines its orbital angular momentum and describes the shape of the orbital of a particle within the associated geometry. It is worth noting that the orbitals can take even more complex shapes according to the higher values of $l$. A spherical orbit ($l = 0$) can be oriented in space in only one way. However, an orbital that has polar ($l = 1$) or cloverleaf ($l = 2$) shape can point in different directions. That is why, in order to describe the orientation in space of a particular orbital, one always needs the magnetic quantum number $m$.

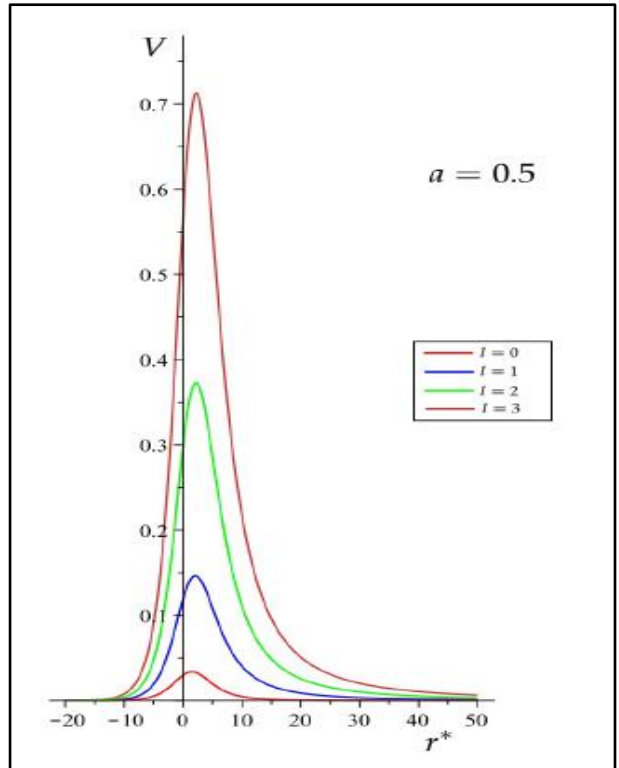

Figure 1: Plot of $V$ versus $r^*$. The physical parameters are chosen to be with $M = 1$ and $a = 0.5$. As $l$ gets bigger values, the potential barrier between the horizon and spatial infinity exhaustively increases

Figure 1 exhibits the plot of $V(r)$ versus $r^*$ for various values of the azimuthal quantum number $l$ with $M = 1$ and $a = 0.5$. It can be deduced from Figure 1 that when $l$-parameter takes higher values; the effective potential tends to make a BP at a specific point which is among the event horizon and spatial infinity. This means that the scalar waves having very high azimuthal quantum number ($l \gg 1$) that are not sufficiently energetic, could not pass that BP and would be confined in a small region. To obtain those RMs, as being discussed in the following section, we shall make the analysis around the NH region of the GHSBH.

### 2.3.2 RMFs and Entropy/Area Spectra of GHSBH

In principle, Eq. (21) is solved for the QNMs with a set of boundary conditions: purely ingoing wave at the horizon and purely outgoing waves at the infinity. But unfortunately, Eq. (21) cannot be solved, analytically. Therefore, to overcome this difficulty, one should take the help of some approximate method. In this section, we shall follow a method prescribed in [45,48,51,114], which considers the wave dynamics in the vicinity of the event horizon. Since the effective potential (27) vanishes at the horizon ($r^* \to -\infty$) and makes a barrier for the scalar waves with high azimuthal quantum numbers ($l \gg 1$) in the intermediate region, the RMs are defined to be those for which one has purely ingoing plane wave at the horizon and no wave at BP's location: the latter condition is trivially satisfied. Namely, the relevant RMs should satisfy

$$H(r)|_{RM} \sim \begin{cases} e^{i\omega r^*} & \text{at} \quad r^* \to -\infty \\ 0 & \text{at} \quad \text{BP} \end{cases}. \tag{72}$$

The series expansion of the metric function around the event horizon is given as

$$f(r) \simeq f'(r_+)(r - r_+) + \frac{f''(r_+)}{2}(r - r_+)^2 + O[(r - r_+)^3]$$

$$= 2\kappa y + +\frac{f''(r_+)}{2}y^2 + O(y^3), \tag{73}$$

with $y = r - r_+$. With this new variable, we apply the Taylor expansion around $y = 0$ to Eq. (27) and obtain the NH form of the effective potential as

$$V(y) \simeq 2\kappa y[l(l+1)(C+Dy) + 2\kappa(G+Hy) - 2\kappa yN], \tag{74}$$

while the parameters seen in the above potential are given by

$$C = \frac{1}{z}, \quad D = -\frac{2x}{z^2}, \quad G = \frac{x}{z}, \quad H = -\frac{x^2+a^2}{z^2}, \quad N = \frac{a^2}{z^2}, \tag{75}$$

and

$$x = r_+ - a, \quad z = r_+(r_+ - 2a). \tag{76}$$

Around the event horizon, the tortoise coordinate is further expressed as follows:

$$r^* \simeq \frac{1}{2\kappa} \ln y. \tag{77}$$

Thus, in the NH region, the ZE (21) behaves as follows

$$-4\kappa^2 y^2 \frac{d^2 H(y)}{dy^2} - 4\kappa^2 y \frac{dH(y)}{dy} + [V(y) - \omega^2]H(y) = 0. \tag{78}$$

which is a differential equation with two separate solutions. The solutions can be expressed in terms of the Whittaker functions [113]. The solution can be transformed to the CH functions [113] and thus we have

$$H(y) \sim C_1 y^{\frac{i\omega}{2\kappa}} M(\tilde{a}, \tilde{b}, \tilde{c}y) + C_2 y^{\frac{i\omega}{2\kappa}} U(\tilde{a}, \tilde{b}, \tilde{c}y). \tag{79}$$

The parameters of the above function are given by

$$\tilde{a} = -i\frac{\gamma}{\lambda\sqrt{\kappa}} + \frac{\tilde{b}}{2}, \quad \tilde{b} = 1 + i\frac{\omega}{\kappa}, \quad \tilde{c} = i\frac{\lambda}{2z\sqrt{\kappa}}, \tag{80}$$

with

$$\gamma = 2\kappa x + l(l+1), \quad \lambda = 4\sqrt{xl(l+1) + \kappa(z + 3a^2)}. \tag{82}$$

By using one of the transformations of the CH functions [113], we are enabled to obtain the NH ($y \ll 1$) form of the solution (79) as

$$H(y) \sim \left[C_1 + C_2 \frac{\Gamma(1-\tilde{b})}{\Gamma(1+\tilde{a}-\tilde{b})}\right] y^{\frac{i\omega}{2\kappa}} + C_2 \frac{\Gamma(\tilde{b}-1)}{\Gamma(\tilde{a})} y^{-\frac{i\omega}{2\kappa}}, \tag{83}$$

Since the RMs impose vanishing outgoing waves at the horizon, the second term must be terminated with the poles of the Gamma function seen in the denominator; i.e., $\tilde{a} = -n$ with $n = 0, 1, 2, .....$. Thence, we read the RMFs of the GHSBH as

$$\omega_n = \frac{\sqrt{\kappa}[2\kappa x + l(l+1)]}{2\sqrt{xl(l+1) + \kappa(x^2 + 2a^2)}} + i(2n+1)\kappa$$

$$\simeq \frac{\sqrt{\kappa}}{2\sqrt{x}} l + i(2n+1)\kappa, \qquad l \gg 1. \tag{84}$$

The transition frequency for the highly excited states (when the resonance parameter [144] $n \to \infty$ and therefore $\omega_I \gg \omega_R$), reads

$$\Delta\omega \approx \Delta\omega_I = 2\kappa = \frac{4\pi T_H}{\hbar}. \tag{85}$$

Substituting this transition frequency into Eq. (3), we have

$$I_{adb} = \frac{\hbar}{4\pi} S^{BH}. \tag{86}$$

Recalling the BSQR ($I_{adb} = \hbar n$) [27], the entropy spectrum reads

$$S_n^{BH} = 4\pi n. \tag{87}$$

Knowing that $S^{BH} = \mathcal{A}/4\hbar$, one can also find the area spectrum as

$$\mathcal{A}_n = 16\pi\hbar n, \tag{88}$$

and, accordingly the minimum area spacing reads

$$\Delta\mathcal{A}_{min} = 16\pi\hbar. \tag{89}$$

The results obtained for the entropy/area spectra of the GHSBH from the RMs show that the spectra are evenly spaced. The same conclusion was obtained in the studies of [45,48,51,114], although the value of $\varepsilon$ is now twice the one obtained from the BQNMs. In any case, our findings are in accordance with the conjecture of Kothawala et al. [42] in which it is argued that the BHs of Einstein's theories should have equidistant area spectrum.

# Chapter 3

# SPECTROSCOPY OF THE ZZLBH[2]

## 3.1 ZZLBH Spacetime

In this section, we introduce the four-dimensional Lifshitz spacetimes and its special case, that is, ZZLBH [117]. Conformal gravity (CG) covers gravity theories that are invariant under Weyl transformations. CG is adapted to static and asymptotically Lifshitz BH solutions, has received intensive attention from the researchers studying condensed matter and quantum field theories [120]. The Lifshitz BHs are invariant under anisotropic scale and characterize the gravitational dual of strange metals [121].

The action of the Einstein-Weyl gravity [117] is given by

$$S = \frac{1}{2\tilde{\kappa}^2} \int \sqrt{-g} d^4x \left( R - 2\Lambda + \frac{1}{2}\alpha |Weyl|^2 \right), \qquad (90)$$

where $\tilde{\kappa}^2 = 8\pi$, $|Weyl|^2 = R^{\mu\nu\rho\sigma}R_{\mu\nu\rho\sigma} - 2R^{\mu\nu}R_{\mu\nu} + \frac{1}{3}R^2$, and $\alpha = \frac{z^2+2z+3}{4z(z-4)}$ (constant), which diverges ($\alpha = \infty$) with $z = 0$ and/or $z = 4$. The Lifshitz BH solutions exist in the CG theory for both $z = 0$ and $z = 4$ [117,150]; however, when $z = 3$ and/or $z = 4$, the Lifshitz BHs appear in the Hořava-Lifshitz gravity [117,118,151].

Here, we especially focus on the $z = 0$ solution, namely ZZLBH of the CG theory whose metric is given by [117]

---

[2] This chapter is mainly quoted from [155,157].

$$ds^2 = -f(r)dt^2 + \frac{4dr^2}{r^2 f(r)} + r^2 d\Omega_{2,k}^2, \qquad (91)$$

where $f(r)$ is the metric function defined by

$$f(r) = 1 + \frac{c}{r^2} + \frac{c^2 - k^2}{3r^4}. \qquad (92)$$

The above metric is conformal to (A)dS (AdS if $c + 2k < 1$ and dS if $c + 2k > 1$) with [117,151]

$$d\Omega_{2,k}^2 = \begin{cases} d\theta^2 + d\phi^2, & k = 0, \\ d\theta^2 + \sin^2\theta \, d\phi^2, & k = 1, \\ d\theta^2 + \sinh^2\theta \, d\phi^2, & k = -1. \end{cases} \qquad (93)$$

The metric solution has a curvature singularity at $r = 0$, which becomes naked for $k = 0$, and an event horizon $r_+$ for $k = \pm 1$ solution, which is expressed as [117,151]:

$$r_+^2 = \frac{1}{6}\left(\sqrt{3(4 - c^2)} - 3c\right). \qquad (94)$$

Notice that the requirement of $r_+^2 \geq 0$ is bound on the inequality: $-2 \leq c < 1$. The solution becomes extremal when $c = -2$. For simplicity, throughout this chapter, we regard the choice of $c^2 = k^2 = 1$ for which the solution becomes a dS BH. Thus, in accordance with our choice, the metric and the metric function become, respectively:

$$ds^2 = -f(r)dt^2 + \frac{4}{r^2 f(r)} dr^2 + r^2[d\theta^2 + \sin^2\theta \, d\phi^2], \qquad (95)$$

with

$$f(r) = 1 - \frac{r_+^2}{r^2}. \qquad (96)$$

Thus, at spatial infinity, the Ricci and Kretschmann scalars of the ZZLBH can be found as follows

$$R = R_\lambda^\lambda \sim \frac{5(c^4 - 2c^2 + 1)}{12 r^8}, \qquad (97)$$

$$K = R^{\mu\nu\rho\sigma} R_{\mu\nu\rho\sigma} \sim \frac{25(c^4 - 2c^2 + 1)}{12 r^8}. \qquad (98)$$

On the other hand, the quasilocal mass $M_{QL}$ of the ZZLBH can be obtained from Wald's entropy formula [142,152] by integrating the first law of thermodynamics $dM_{QL} =$ $T_H dS^{BH}$. Also, one can check this result with the quasilocal mass computation via Brown-York (BY) formalism [153,154]. In fact, the mass calculations of the ZZLBH were done in [140] and the result obtained was

$$4M_{QL} = 4M_{BY} = r_+^2. \tag{99}$$

By performing the surface gravity calculation [4,5,142], we obtain

$$\kappa = \frac{1}{2}. \tag{100}$$

Accordingly, the Hawking or BH temperature [142] of the ZZLBH reads

$$T_H = \frac{\hbar \kappa}{2\pi} = \frac{\hbar}{4\pi}. \tag{101}$$

As it can be deduced from the constant Hawking temperature, the ZZLBH radiates with isothermal process. After substituting $\kappa$ into $dM = T_H dS^{BH}$, we derive the entropy as

$$S^{BH} = \frac{\mathcal{A}}{4\hbar} = \frac{\pi r_+^2}{\hbar}. \tag{102}$$

## 3.2 Quantization of the ZZLBH with Scalar Waves

In this section, we study the entropy/area spectra of a perturbed ZZLBH via MM [16]. To this end, we compute the QNMs of the ZZLBH [155].

### 3.2.1 Separation of the Massive KGE on the ZZLBH

The QNMs of the considered BH can be extracted by solving the KGE with the proper physical boundary conditions. The event horizon boundary condition suggests that only ingoing waves carry the QNMs at the event horizon (i.e., no outgoing waves are allowed at the event horizon) and the boundary condition at spatial infinity implies that the only waves allowed to survive at spatial infinity are the outgoing waves. The latter boundary condition is suitable for the effective potentials of bumpy shape which terminates at its both ends. However, as seen in Figure 2 and can be deduced from Eqs. (108) and (109), the potential diverges for very massive ($m \to \infty$) scalar particles, it never terminates at spatial infinity. Thus, the potential blocks the waves that come off from the BH and prevents them to reach to spatial infinity. For this reason, we will consider very massive scalar particles and utilize the particular method given in [45,51,114-116], in which only the QNMs have purely ingoing plane waves at the horizon and no waves at spatial infinity. In other words, we will use the NH boundary condition in order for finding the QNMs of the ZZLBH.

To this end, first we consider the massive KGE:

$$\frac{1}{\sqrt{-g}}\partial_\mu\left(\sqrt{-g}g^{\mu\nu}\partial_\nu\right)\Psi - m^2\Psi = 0, \tag{103}$$

where $\Psi$ adopts the ansatz for the above wave equation chosen as

$$\Psi = \frac{1}{r}F(r)e^{i\omega t}Y_{lm}(\theta,\varphi), \tag{104}$$

in which $F(r)$ is the function of $r$, while $Y_{lm}(\theta,\varphi)$ represents the spherical harmonics with the eigenvalues $-l(l+1)$ and $m$. After a straightforward calculation, separation of the KGE yields a radial part which later reduces to a SLE [61]:

$$\left[-\frac{d}{dr^{*2}} + V_{eff}(r) - \omega^2\right]F(r) = 0, \tag{105}$$

where $r^*$ and $V_{eff}(r)$ stand for the tortoise coordinate and the effective potential, respectively. The tortoise coordinate is given by the following integral

$$r^* = 2\int\frac{dr}{rf(r)}, \tag{106}$$

which results in

$$r^* = \ln\left(\frac{r^2}{r_+^2} - 1\right). \tag{107}$$

One may check that the limits of $r^*$ admit the following:

$$\lim_{r\to r_+} r^* = -\infty, \tag{108}$$

$$\lim_{r\to\infty} r^* = \infty. \tag{109}$$

The effective potential in Eq. (105) is found to be as follows

$$V_{eff}(r) = f(r)\left\{\frac{l(l+1)}{r^2} + \frac{1}{4}\left[f(r) + r\frac{df(r)}{dr}\right] + m^2\right\}, \qquad (110)$$

whose limits are:

$$\lim_{r \to r_+} V_{eff}(r) = 0, \qquad (111)$$

$$\lim_{r \to \infty} V_{eff}(r) = m^2 + \frac{1}{4}. \qquad (112)$$

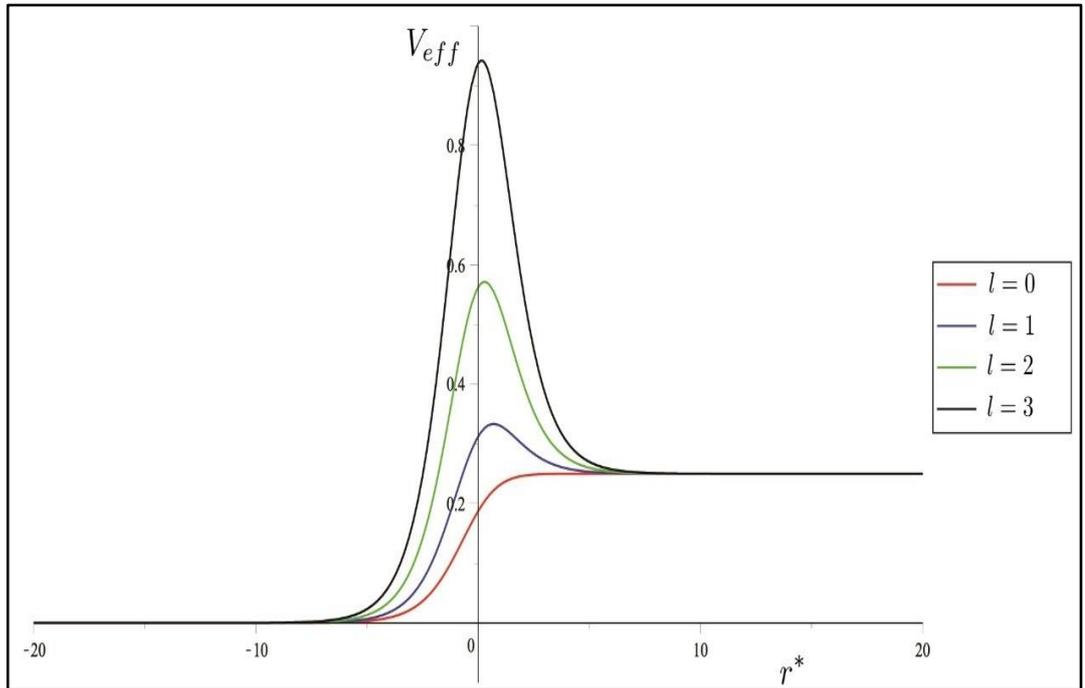

Figure 2: Effective potential versus tortoise coordinate graph for various orbital quantum numbers

It is obvious from Figure 2 that the potential is finite at infinite radius for any (constant) $m$. The potential diverges when the scalar particle is very massive. Namely, the waves tend to cease as $m \to \infty$.

### 3.2.2 QNMFs and Entropy/Area Spectra of the ZZLBH

In this section, we nudge the ZZLBH by the massive scalar fields propagating in the NH region and obtain their corresponding QNMFs.

One can expand the metric function of the ZZLBH to the following series around $r_+$ in terms of the surface gravity $\kappa$:

$$f = f(r_+) + f'(r_+)(r - r_+) + O[(r - r_+)^2] \approx 2\kappa y, \qquad (113)$$

where $y = r - r_h$ and prime (′) denotes the derivative with respect to $r$. Substituting Eq. (113) into Eq. (110) and performing Taylor expansion, we obtain the NH form of the effective potential as

$$V_{NH}(y) = 4\kappa G y[G^2(1 - 2Gy)l(l + 1) + \kappa(1 + 2Gy) + m^2], \qquad (114)$$

with the parameter $G = \frac{1}{r_+}$. The tortoise coordinate reads $r^* \simeq \frac{1}{2\kappa} \ln y$ in the NH region, which allows us to find the NH SLE:

$$\left[-4\kappa^2 y \left(y \frac{d^2}{dy^2} + \frac{d}{dy}\right) + V_{NH}(y) - \omega^2\right] F(y) = 0. \qquad (115)$$

The solutions of the above equation can be written in terms of the CH functions of the first and second kinds [113] as

$$F(y) = y^{\frac{i\omega}{2\kappa}} e^{-\frac{z}{2}} [C_1 M(a, b, z) + C_2 U(a, b, z)], \qquad (116)$$

where the parameters read

$$a = \frac{\lambda}{\sqrt{\delta}} + \frac{b}{2}, \quad b = 1 + i\frac{\omega}{\kappa}, \quad z = 2iG\sqrt{\delta} y, \qquad (117)$$

with

$$\lambda = \frac{1}{2}\left\{1 + \frac{1}{\kappa}[l(l+1)G^2 + m^2]\right\}, \quad \delta = 2\left[l(l+1)\frac{G^2}{\kappa} - 1\right]. \qquad (118)$$

By using the limiting forms of the CH functions [113], we find the NH limit of the solution (116) as

$$F(y) \sim \left[C_1 + C_2 \frac{\Gamma(1-b)}{\Gamma(1+a-b)}\right] y^{\frac{i\omega}{2\kappa}} + C_2 \frac{\Gamma(b-1)}{\Gamma(a)} y^{-\frac{i\omega}{2\kappa}}. \qquad (119)$$

Alternatively, we can express Eq. (119) in terms of $r^*$ ($y \simeq e^{2\kappa r^*}$), hence distinguish the NH ingoing and outgoing waves:

$$\Psi \sim \left[C_1 + C_2 \frac{\Gamma(1-b)}{\Gamma(1+a-b)}\right] e^{i\omega(t+r^*)} + C_2 \frac{\Gamma(b-1)}{\Gamma(a)} e^{i\omega(t-r^*)}. \qquad (120)$$

Imposing the boundary conditions for the QNMs, i.e.,

$$F(r^*)_{QNM} \sim \begin{cases} e^{i\omega r^*} & \text{at} \quad r^* \to -\infty \\ 0 & \text{at} \quad r^* \to \infty, \end{cases} \quad (121)$$

the solution with the coefficient $C_2$ should vanish. By using the pole structure of the Gamma function, the outgoing waves are terminated for $a = -n$ with $n = 0, 1, 2, ...$. This yields the QNMFs of the ZZLBH as

$$\omega_n = 2\kappa \left[ \left(n + \tfrac{1}{2}\right)i + \tfrac{\lambda}{\sqrt{\delta}} \right], \quad (122)$$

where $n$ is known as the overtone quantum number [144]. Accordingly, the transition frequency between two highly excited subsequent states ($\omega_I \gg \omega_R$) is easily obtained as follows

$$\Delta\omega \approx \Delta\omega_I = 2\kappa = \frac{4\pi T_H}{\hbar}. \quad (123)$$

Subsequently, using the adiabatic invariant quantity (3) and BSQR

$$I_{adb} = \frac{\hbar}{4\pi} \int \frac{dM}{T_H} = \frac{\hbar}{4\pi} S^{BH} = \hbar n, \quad (124)$$

we can read the entropy/area spectra of the ZZLBH as follows

$$S_n^{BH} = \frac{\mathcal{A}_n}{4\hbar} = 4\pi n, \quad (125)$$

and

$$\mathcal{A}_n = 16\pi\hbar n. \quad (126)$$

Therefore, the minimum spacing of the BH area becomes

$$\Delta\mathcal{A}_{min} = 16\pi\hbar, \quad (127)$$

which agrees with the Bekenstein's conjecture [10], and the equispacing of the entropy/area spectra of the ZZLBH support the Kothawala et al.'s hypothesis [42].

## 3.3 Quantization of the ZZLBH with Fermionic Waves

In GR, to study the Dirac equation in curved spacetimes, many formalisms have been developed such as NP and spinor formalisms [156]. In this section, in order to solve

Dirac equation in the ZZLBH geometry, we use NP formalism [156] and consider a massive Dirac field. Then, we obtain the Dirac BQNMs of the ZZLBH by using the method of [107]. Finally, we derive the spectroscopy of the ZZLBH from the Dirac BQNMs [157].

### 3.3.1 Separation of the CDEs on the ZZLBH

In this section, we separate the Dirac equation and give the solution of angular equation with the separation constant $\lambda$. Then, we show how the radial equation reduces to the ZE [61] with an effective potential in the NH.

The basis vectors of the null tetrad [156] in the geometry defined by the opposite sign of the line element (95) are chosen as

$$l_\mu = \sqrt{\frac{f(r)}{2}}\left[1, -\frac{2}{rf(r)}, 0, 0\right],$$

$$n_\mu = \sqrt{\frac{f(r)}{2}}\left[1, \frac{2}{rf(r)}, 0, 0\right],$$

$$m_\mu = -\frac{r}{\sqrt{2}}[0, 0, 1, i\sin(\theta)],$$

$$\overline{m}_\mu = \frac{r}{\sqrt{2}}[0, 0, -1, i\sin(\theta)]. \tag{128}$$

The non-zero NP spin coefficients [156] regarding the above covariant null tetrad are found as

$$\alpha = -\beta = -\frac{\sqrt{2}}{4}\frac{\cot(\theta)}{r},$$

$$\varepsilon = \gamma = \frac{\sqrt{2}}{16}\frac{r}{\sqrt{f(r)}}[\partial_r f(r)],$$

$$\rho = \mu = -\frac{\sqrt{2}}{4}\sqrt{f(r)}. \tag{129}$$

In NP formalism, in the case of $q = 0$ test Dirac particles, the Dirac equations can be expressed by the well-known Chandrasekhar-Dirac equations (CDEs) [61] with the aid of the spin coefficients as follows

$$(D + \varepsilon - \rho)F_1 + (\bar{\delta} + \pi - \alpha)F_2 = i\mu_p G_1,$$

$$(\Delta + \mu - \gamma)F_2 + (\delta + \beta - \tau)F_1 = i\mu_p G_2,$$

$$(D + \bar{\varepsilon} - \bar{\rho})G_2 - (\delta + \bar{\pi} - \bar{\alpha})G_1 = i\mu_p F_2,$$

$$(\Delta + \bar{\mu} - \bar{\gamma})G_1 - (\bar{\delta} + \bar{\beta} - \bar{\tau})G_2 = i\mu_p F_1, \tag{130}$$

where $\mu^* = \sqrt{2}\mu_p$ is the mass of the uncharged Dirac particles. The directional derivatives corresponding to the null tetrad are defined as [61]

$$D = l^j \partial_j, \quad \Delta = n^j \partial_j, \quad \delta = m^j \partial_j, \quad \bar{\delta} = \bar{m}^j \partial_j, \tag{131}$$

where a bar ($^-$) over a quantity denotes complex conjugation. The wave functions $F_1$, $F_2$, $G_1$, and $G_2$ which represent the Dirac spinors are assumed to be [61]

$$F_1 = f_1(r)A_1(\theta)\exp[i(\omega t + m\phi)],$$

$$G_1 = g_1(r)A_2(\theta)\exp[i(\omega t + m\phi)],$$

$$F_2 = f_2(r)A_3(\theta)\exp[i(\omega t + m\phi)],$$

$$G_2 = g_2(r)A_4(\theta)\exp[i(\omega t + m\phi)], \tag{132}$$

where $\omega$ (positive and real) and $m$ are the frequency of the incoming wave related to the energy of the Dirac particle and the azimuthal quantum number of the wave, respectively.

Substituting Eqs. (129), (131) and (132) into Eq. (130), one can simply get

$$\frac{\tilde{Z}f_1(r)}{f_2(r)} + \frac{LA_3(\theta)}{A_1(\theta)} - i\mu^* r \frac{g_1(r)A_2(\theta)}{f_2(r)A_1(\theta)} = 0,$$

$$\frac{\tilde{Z}f_1(r)}{f_2(r)} + \frac{LA_3(\theta)}{A_1(\theta)} - i\mu^* r \frac{g_1(r)A_2(\theta)}{f_2(r)A_1(\theta)} = 0,$$

$$\frac{\tilde{Z}g_2(r)}{g_1(r)} - \frac{L^\dagger A_2(\theta)}{A_4(\theta)} - i\mu^* r \frac{f_2(r)A_3(\theta)}{g_1(r)A_4(\theta)} = 0,$$

$$\frac{\bar{\tilde{Z}}g_1(r)}{g_2(r)} + \frac{LA_4(\theta)}{A_2(\theta)} + i\mu^* r \frac{f_1(r)A_1(\theta)}{g_2(r)A_2(\theta)} = 0,$$

(133)

with the radial and the angular operators, respectively

$$\tilde{Z} = \frac{i\omega r}{\sqrt{f(r)}} + \frac{1}{2}r^2\sqrt{f(r)}\partial_r + \frac{1}{8}\frac{r^2}{\sqrt{f(r)}}\left[\partial_r\sqrt{f(r)}\right] + \frac{1}{2}r\sqrt{f(r)},$$

$$\bar{\tilde{Z}} = -\frac{i\omega r}{\sqrt{f(r)}} + \frac{1}{2}r^2\sqrt{f(r)}\partial_r + \frac{1}{8}\frac{r^2}{\sqrt{f(r)}}\left[\partial_r\sqrt{f(r)}\right] + \frac{1}{2}r\sqrt{f(r)},$$ (134)

and

$$L = \partial_\theta + \frac{m}{\sin\theta} + \frac{\cot\theta}{2},$$

$$L^\dagger = \partial_\theta - \frac{m}{\sin\theta} + \frac{\cot\theta}{2}.$$ (135)

As it is obvious from Eq. (133) that $\{f_1, f_2, g_1, g_2\}$ and $\{A_1, A_2, A_3, A_4\}$ are the functions of two distinct variables $r$ and $\theta$, respectively; one can introduce a separation constant $\lambda$ and assume that

$$f_1(r) = g_2(r),$$
$$f_2(r) = g_1(r),$$
$$A_1(\theta) = A_2(\theta),$$
$$A_3(\theta) = A_4(\theta).$$ (136)

to split Eq. (133) into two sets of radial and angular equations

$$\bar{\tilde{Z}}g_1(z) = (\lambda - i\mu^* r)g_2(z),$$
$$\tilde{Z}g_2(z) = (\lambda + i\mu^* r)g_1(z),$$ (137)

and

$$L^\dagger A_1(\theta) = \lambda A_3(\theta),$$
$$LA_3(\theta) = -\lambda A_1(\theta).$$ (138)

At this point it would me more convenient to make the choice of massless Dirac particles ($\mu^* = 0$) to deal with above sets of equations.

In the spherical case, the angular operators (or the so-called laddering operators) $L^\dagger$ and $L$ lead the spin weighted spheroidal harmonics $_sY_{l'}^m(\theta)$ [156-161] and they are governed by

$$\left(\partial_\theta - \frac{m}{\sin\theta} - s\cot\theta\right) {}_sY_{l'}^m(\theta) = -\sqrt{(l'-s)(l'+s+1)}\, {}_{s+1}Y_{l'}^m(\theta),$$

$$\left(\partial_\theta + \frac{m}{\sin\theta} + s\cot\theta\right) {}_sY_{l'}^m(\theta) = \sqrt{(l'+s)(l'-s+1)}\, {}_{s-1}Y_{l'}^m(\theta). \quad (139)$$

The spin weighted spheroidal harmonics $_sY_{l'}^m(\theta)$ has the following generic form [161]

$$_sY_{l'}^m(\theta,\phi) = \exp(im\phi)\sqrt{\frac{2l'+1}{4\pi}\frac{(l'+m)!\,(l'-m)!}{(l'+s)!\,(l'-s)!}}\left[\sin\left(\frac{\theta}{2}\right)\right]^{2l'}$$

$$\times \sum_{r=-l'}^{l'}(-1)^{l'+m-r}\binom{l'-s}{r-s}\binom{l'+s}{r-m}\left[\cot\left(\frac{\theta}{2}\right)\right]^{2r-m-s}, \quad (140)$$

where $l'$ is the angular quantum number and $s$ is the spin weight with $l' = |s|, |s|+1, |s|+2, ...$ and $-l' < m < +l'$. Thus, having $s = \pm\frac{1}{2}$, and comparing Eq. (138) with Eq. (139) one can identify

$$A_1(\theta) = {}_{-\frac{1}{2}}Y_{l'}^m(\theta),$$

$$A_3(\theta) = {}_{\frac{1}{2}}Y_{l'}^m(\theta), \quad (141)$$

and obtain the separation constant $\lambda$ which is the eigenvalue of the spin weighted spheroidal harmonic [161] equation as

$$\lambda = -\left(l' + \frac{1}{2}\right). \quad (142)$$

Now, we give the effective potential and obtain the solution to the radial part of the Dirac equation (137). In fact, it is appropriate to alter the radial equations in SLEs with effective potentials, to investigate the BQNMs [109-111].

Defining a new function in terms of the metric function

$$Y(r) = \frac{1}{8}\frac{r^2}{\sqrt{f(r)}}\partial_r f(r) + \frac{1}{2}r\sqrt{f(r)}, \qquad (143)$$

and substituting following scalings

$$g_1(r) = G_1(r)\, exp\left[-2\int \frac{Y(r)}{r^2\sqrt{f(r)}}dr\right],$$

$$g_2(r) = G_2(r)\, exp\left[-2\int \frac{Y(r)}{r^2\sqrt{f(r)}}dr\right], \qquad (144)$$

into the radial equations (137) and setting $\mu^* = 0$, one gets

$$[A(r)\partial_r - i\omega]G_1(r) = B(r)G_2(r),$$

$$[A(r)\partial_r + i\omega]G_2(r) = B(r)G_1(r), \qquad (145)$$

where the functions $A(r)$ and $B(r)$ read

$$A(r) = \frac{rf(r)}{2} \quad \text{and} \quad B(r) = \frac{\lambda\sqrt{f(r)}}{r}, \qquad (146)$$

respectively. Introducing the tortoise coordinate as

$$dr_* = \frac{1}{A(r)}dr, \qquad (147)$$

leads us to express Eq. (145) as follows

$$[\partial_{r_*} - i\omega]G_1(r_*) = B(r)G_2(r_*),$$

$$[\partial_{r_*} + i\omega]G_2(r_*) = B(r)G_1(r_*). \qquad (148)$$

To decouple the above equations, we consider the solutions of the form:

$$G_1(r_*) = P_1(r_*) + P_2(r_*),$$

$$G_2(r_*) = P_1(r_*) - P_2(r_*), \qquad (149)$$

which yields the two radial equations (SLEs or the so-called ZEs [61])

$$\partial_{r_*}^2 P_1(r_*) + [\omega^2 - V_1]P_1(r_*) = 0,$$

$$\partial_{r_*}^2 P_2(r_*) + [\omega^2 - V_2]P_2(r_*) = 0, \qquad (150)$$

whose associated Zerilli potentials are given by

$$V_1 = B^2(r) + \partial_{r_*} B(r) = \frac{\lambda\sqrt{r^2 - r_+^2}}{2r^4}\left(2\lambda\sqrt{r^2 - r_+^2} + 2r_+^2 - r^2\right),$$

$$V_2 = B^2(r) - \partial_{r_*} B(r) = \frac{\lambda\sqrt{r^2-r_+^2}}{2r^4}\left(2\lambda\sqrt{r^2 - r_+^2} - 2r_+^2 + r^2\right). \tag{151}$$

### 3.3.2 BQNMFs and Entropy/Area Spectra of the ZZLBH

In this section, our interest is to compute the fermion BQNMs. First, we impose the purely ingoing wave condition at the event horizon for QNMs to appear [61]. Then, we impose the DBC and the NBC for getting the resonance conditions. At this point, we use an iteration method in order to derive the RMs [106,107].

Letting $r = xr_+ + r_+$ in $A(r)$ gives

$$A(x) = r_+ x \frac{(x+2)}{2(x+1)}, \tag{152}$$

with its NH form:

$$A_{NH}(x) \approx r_+ x + O(x^2) = 2\kappa r_+ x. \tag{153}$$

This allows us to write the NH forms of the Zerilli potentials (151) as follows:

$$V_1^{NH}(x) = \frac{\lambda}{\sqrt{2}r_+}\sqrt{x} + \frac{2\lambda^2}{r_+^2}x + O(x^{3/2}),$$

$$V_2^{NH}(x) = -\frac{\lambda}{\sqrt{2}r_+}\sqrt{x} + \frac{2\lambda^2}{r_+^2}x + O(x^{3/2}). \tag{154}$$

Setting a new coordinate

$$y = \int \frac{\kappa r_+}{A_{NH}(x)} dx = \frac{1}{2}\ln(x) = \kappa r_*, \tag{155}$$

whose limits are $\lim_{r \to r_+} y = -\infty$ and $\lim_{r \to \infty} y = \infty$; one can express the radial coordinate $x$ in terms of the surface gravity as

$$x = exp(2y) = exp(2\kappa r_*). \tag{156}$$

Therefore, we can recast the NH effective potentials (154), in the leading order terms as follows

$$V_1^{NH}(y) = \frac{\lambda}{\sqrt{2}r_+} exp(y),$$

$$V_2^{NH}(y) = -\frac{\lambda}{\sqrt{2}r_+} exp(y). \qquad (157)$$

The NH ZEs, in which $\widetilde{\omega} = \omega/\kappa$, are found as

$$\frac{d^2}{dy^2}P_1(y) + \left[\widetilde{\omega}^2 - \frac{V_1^{NH}(y)}{\kappa^2}\right]P_1(y) = 0,$$

$$\frac{d^2}{dy^2}P_2(y) + \left[\widetilde{\omega}^2 - \frac{V_2^{NH}(y)}{\kappa^2}\right]P_2(y) = 0, \qquad (158)$$

These can be solved after inserting the value of the separation constant. The solutions of these equations are obtained in terms of the Bessel functions of the first and the second kind [113]:

$$P_j(y) = C_{j1}J_{-2i\widetilde{\omega}}\left[\aleph 2^{5/4}\sqrt{\frac{2l'+1}{r_+}}exp\left(\frac{y}{2}\right)\right] + C_{j2}Y_{-2i\widetilde{\omega}}\left[\aleph 2^{5/4}\sqrt{\frac{2l'+1}{r_+}}exp\left(\frac{y}{2}\right)\right], \qquad (159)$$

where $C_{j1}$ and $C_{j2}$ are constants and

$$\aleph = (i)^{j-1}, \quad j = 1,2. \qquad (160)$$

The solution of the NH ZEs (158) can be rewritten in a compact form as follows

$$P_j(x) = C_{j1}J_{-2i\widetilde{\omega}}\left(4\aleph\sqrt{\Omega\sqrt{x}}\right) + C_{j2}Y_{-2i\widetilde{\omega}}\left(4\aleph\sqrt{\Omega\sqrt{x}}\right), \qquad (161)$$

with the parameter

$$\Omega = \frac{2l'+1}{2\sqrt{2}r_+}. \qquad (162)$$

Using Eqs. (42) and (43), one can obtain the NH ($e^{y/2} \ll 1$) form of the solution as

$$P_j \sim C_{j1}\frac{(2\aleph\sqrt{\Omega})^{-2i\widetilde{\omega}}}{\Gamma(1-2i\widetilde{\omega})}exp(-i\widetilde{\omega}y) - C_{j2}\frac{1}{\pi}\Gamma(-2i\widetilde{\omega})(2\aleph\sqrt{\Omega})^{2i\widetilde{\omega}}exp(i\widetilde{\omega}y)$$

$$= C_{j1}\frac{(2\aleph\sqrt{\Omega})^{-2i\widetilde{\omega}}}{\Gamma(1-2i\widetilde{\omega})}exp(-i\omega r_*) - C_{j2}\frac{1}{\pi}\Gamma(-2i\widetilde{\omega})(2\aleph\sqrt{\Omega})^{2i\widetilde{\omega}}exp(i\omega r_*) \qquad (163)$$

Imposing the boundary condition at the event horizon for BQNMs requires us to vanish the outgoing waves by choosing $C_{j2} = 0$. Therefore, the proper solution of Eq. (158) becomes

$$P_j(x) = C_{j1} J_{-2i\widetilde{\omega}}\left(4\aleph\sqrt{\Omega\sqrt{x}}\right). \tag{164}$$

Taking the DBC into account at the confining cage $x_m$ [52,106,107,112]

$$P_j(x)|_{x=x_m} = 0, \tag{165}$$

one gets the condition

$$J_{-2i\widetilde{\omega}}\left(4\aleph\sqrt{\Omega\sqrt{x_m}}\right) = 0. \tag{166}$$

With the aid of the relation given in Eq. (48), the boundary condition (166) can be stated as

$$tan(2i\widetilde{\omega}\pi) = \frac{J_{2i\widetilde{\omega}}\left(4\aleph\sqrt{\Omega\sqrt{x_m}}\right)}{Y_{2i\widetilde{\omega}}\left(4\aleph\sqrt{\Omega\sqrt{x_m}}\right)}. \tag{167}$$

The boundary of the cage is at the event horizon. Thus, one can use the limiting forms of the Bessel functions [113] in the above condition and obtain the resonance condition as follows

$$tan(2i\widetilde{\omega}\pi) \sim -\frac{\pi\left(2\aleph\sqrt{\sqrt{z_m}}\right)^{4i\widetilde{\omega}}}{\Gamma(2i\widetilde{\omega})\Gamma(2i\widetilde{\omega}+1)} = i\frac{\pi(\aleph^2)^{2i\widetilde{\omega}}}{2\widetilde{\omega}\Gamma^2(2i\widetilde{\omega})}\left(4\sqrt{z_m}\right)^{2i\widetilde{\omega}}$$

$$= i\frac{\pi exp(-4\pi\widetilde{\omega}/j)}{2\widetilde{\omega}\Gamma^2(2i\widetilde{\omega})}\left(4\sqrt{z_m}\right)^{2i\widetilde{\omega}} \tag{168}$$

where $z_m = \Omega^2 x_m$. Imposing the NBC [52,106,107,112] given by

$$\frac{dP_j(x)}{dx}|_{x=x_m} = 0, \tag{169}$$

we find

$$J_{-2i\widetilde{\omega}-1}\left(4\aleph\sqrt{\sqrt{z_m}}\right) - J_{-2i\widetilde{\omega}+1}\left(4\aleph\sqrt{\sqrt{z_m}}\right) = 0. \tag{170}$$

We combine Eqs. (54) and (170) and express the NBC's resonance condition as

$$tan(2i\widetilde{\omega}\pi) = \frac{J_{2i\widetilde{\omega}}\left(4\aleph\sqrt{\sqrt{z_m}}\right)}{Y_{2i\widetilde{\omega}}\left(4\aleph\sqrt{\sqrt{z_m}}\right)} \left[\frac{-1+J_{2i\widetilde{\omega}+1}\left(4\aleph\sqrt{\sqrt{z_m}}\right)/J_{2i\widetilde{\omega}-1}\left(4\aleph\sqrt{\sqrt{z_m}}\right)}{1-Y_{2i\widetilde{\omega}-1}\left(4\aleph\sqrt{\sqrt{z_m}}\right)/Y_{2i\widetilde{\omega}+1}\left(4\aleph\sqrt{\sqrt{z_m}}\right)}\right]. \tag{171}$$

Using the limiting forms (42) and (43), one can find

$$\frac{J_{2i\widetilde{\omega}+1}\left(4\aleph\sqrt{\sqrt{z_m}}\right)}{J_{2i\widetilde{\omega}-1}\left(4\aleph\sqrt{\sqrt{z_m}}\right)} \equiv \frac{Y_{2i\widetilde{\omega}-1}\left(4\aleph\sqrt{\sqrt{z_m}}\right)}{Y_{2i\widetilde{\omega}+1}\left(4\aleph\sqrt{\sqrt{z_m}}\right)} \sim O(z_m), \tag{172}$$

and read the resonance condition (171) in the NH as

$$tan(2i\widetilde{\omega}\pi) \sim -\frac{J_{2i\widetilde{\omega}-1}\left(4\aleph\sqrt{\sqrt{z_m}}\right)}{Y_{2i\widetilde{\omega}+1}\left(4\aleph\sqrt{\sqrt{z_m}}\right)} = -i\frac{\pi(\aleph^2)^{2i\widetilde{\omega}}}{2\widetilde{\omega}\Gamma^2(2i\widetilde{\omega})}\left(4\sqrt{z_m}\right)^{2i\widetilde{\omega}}$$

$$= -i\frac{\pi exp(-4\pi\widetilde{\omega}/j)}{2\widetilde{\omega}\Gamma^2(2i\widetilde{\omega})}\left(4\sqrt{z_m}\right)^{2i\widetilde{\omega}}. \tag{173}$$

To get the resonance frequencies, we use an iteration method [52,106,107], since the obtained resonances are small quantities. The $0^{th}$ order resonance condition [52,106,107] has the form

$$tan\left(2i\widetilde{\omega}_n^{(0)}\pi\right) = 0, \tag{174}$$

which means that

$$\widetilde{\omega}_n^{(0)} = -i\frac{n}{2}, \quad n = 0, 1, 2, \dots. \tag{175}$$

The $1^{st}$ order resonance condition can be obtained by substituting Eq. (175) into the r.h.s. of (168) and (173), as follows

$$tan\left(2i\widetilde{\omega}_n^{(1)}\pi\right) = \pm i\frac{\pi exp(2i\pi n/j)}{(-in)\Gamma^2(n)}\left(4\sqrt{z_m}\right)^n, \tag{176}$$

which reduces to

$$tan\left(2i\widetilde{\omega}_n^{(1)}\pi\right) = \mp n\frac{\pi}{(n!)^2}\left[(-1)^{2/j}4\sqrt{z_m}\right]^n. \tag{177}$$

Using Eq. (62) yields

$$2i\widetilde{\omega}_n \pi = n\pi \left\{ 1 \mp \frac{1}{(n!)^2} \left[ (-1)^{2/j} 4\sqrt{z_m} \right]^n \right\}. \tag{178}$$

Hence, we find

$$\widetilde{\omega}_n = -i\frac{n}{2} \left\{ 1 \mp \frac{1}{(n!)^2} \left[ (-1)^{2/j} 4\sqrt{z_m} \right]^n \right\}, \tag{179}$$

and read the Dirac BQNMs as follows

$$\omega_n = -i\kappa \frac{n}{2} \left\{ 1 \mp \frac{1}{(n!)^2} \left[ (-1)^{2/j} 4\sqrt{z_m} \right]^n \right\}, \quad n = 0, 1, 2, .... \tag{180}$$

where $n$ stands for the overtone quantum number (resonance parameter) [144].

For the highly excited states ($n \to \infty$), BQNMFs read

$$\omega_n \approx -i\kappa \frac{n}{2}, \tag{181}$$

The transition frequency from MM [16] can be obtained as

$$\Delta\omega_I = \frac{\kappa}{2} = \frac{\pi T_H}{\hbar}. \tag{182}$$

Therefore, the adiabatic invariant quantity (3) [23,24,50] becomes

$$I_{adb} = \frac{\hbar}{\pi} S^{BH}. \tag{183}$$

Using the BSQR [27], the entropy spectrum of ZZLBHs can be determined as

$$S_n^{BH} = \pi n, \tag{184}$$

and using $S^{BH} = \mathcal{A}/4\hbar$, one may obtain the area spectrum

$$\mathcal{A}_n = 4\pi\hbar n, \tag{185}$$

with the minimum spacing given by

$$\Delta \mathcal{A}_{min} = 4\pi\hbar. \tag{186}$$

As such, one concludes that the entropy/area spectra of the ZZLBHs are evenly spaced and are independent from the BH parameters whereas the spacing coefficient reads $\varepsilon = 4\pi$, which is half the Bekenstein's result for the Schwarzschild BH [16,23,24].

# Chapter 4

# SPECTROSCOPY AND SCALAR CLOUDS OF THE RLDBH[3]

## 4.1 RLDBH and MRLDBH Spacetimes

The existence of the RLDBH solutions was discovered by Clément et al. [44] in the EMDA theory whose action is given by [127].

$$S = \frac{1}{16\pi} \int d^4x \sqrt{-g} \left( R - \frac{1}{2} e^{4\phi} \partial_\mu \aleph \partial^\mu \aleph - 2\partial_\mu \phi \partial^\mu \phi - \aleph F_{\mu\nu} \tilde{F}^{\mu\nu} - e^{-2\phi} F^2 \right), \quad (187)$$

where $F_{\mu\nu}$ and $\tilde{F}^{\mu\nu}$ denote the Maxwell tensor (antisymmetric rank-2 tensor field) and its dual, while $R$ is the Ricci scalar. $\varphi$ and $\aleph$ are the dilaton and the axion (pseudoscalar) fields, respectively. The axion field may affirm the existence of the cold dark matter [45,46] in the RLDBH spacetime. This spacetime is described by the following metric [125,127]

$$ds^2 = -f(r)dt^2 + \frac{dr^2}{f(r)} + h(r)\left[d\theta^2 + \sin^2\theta \left(d\varphi - \frac{a}{h(r)}dt\right)^2\right], \quad (188)$$

with the metric functions

$$h(r) = rr_0, \quad (189)$$

$$f(r) = \frac{Z}{h(r)}, \quad (190)$$

where the constant parameter $r_0$ is directly proportional to the background electric charge $Q$: $r_0 = \sqrt{2}Q$. In Eq. (190), $Z = (r - r_2)(r - r_1)$, where, $r_1$ and $r_2$ are two

---

[3] This chapter is mainly quoted from [52,163].

positive roots of the metric function and denote the inner and outer horizons, respectively.

$$r_1 = M - \sqrt{M^2 - a^2}, \tag{191}$$

$$r_2 = M + \sqrt{M^2 - a^2}, \tag{192}$$

where $a$ is the rotation parameter modulating the angular momentum: $J = \frac{ar_0}{2}$, and $M$ is an integration constant related to mass in deriving the RLDBH solution. The NAF structure of the RLDBH geometry entails the $M_{QL}$ computation [153]. As a result, one finds that $M = 2M_{QL}$. The rotation parameter $a$ is related with the angular momentum ($J$) of the RLDBH via $a = \frac{\sqrt{2}J}{Q}$. Meanwhile, it is obvious from Eqs. (191) and (192) that for a BH solution, $M \geq a$. Thus, MRLDBH geometry corresponds to $a = M$.

The background fields of dilaton and axion are given by [125]

$$e^{-2\phi} = \frac{h(r)}{s(r)}, \tag{193}$$

$$\aleph = -\frac{r_0 a \cos\theta}{s(r)}, \tag{194}$$

where $s(r) = r^2 + a^2 \cos^2\theta$. Furthermore, the electromagnetic four-vector potential is given as

$$A_{em} = \frac{1}{\sqrt{2}}(e^{2\phi} dt + a \sin^2\theta \, d\varphi). \tag{195}$$

The Hawking temperature [2,142] of the RLDBH is provided by the surface gravity $\kappa$ definition [142] as

$$T_H = \frac{\hbar\kappa}{2\pi} = \frac{\hbar}{2\pi}\left(\frac{\partial_r f(r)}{2}\bigg|_{r=r_2}\right) = \frac{\hbar(r_2 - r_1)}{4\pi r_2 r_0}. \tag{196}$$

As can be seen from the above equation, in the extreme case $M = a$ (maximal rotation, i.e., $r_1 = 0$), the RLDBH emits radiation with a mass-independent constant temperature (due to the fixed $Q$ value): $T_H = \left(4\sqrt{2}\pi Q\right)^{-1}$ which is nothing but the well-known isothermal process in the subject of the thermodynamics.

The entropy of this BH is given by

$$S^{BH} = \frac{\mathcal{A}}{4\hbar} = \frac{\pi r_2 r_0}{\hbar}. \tag{197}$$

Angular velocity of the RLDBH is given by

$$\Omega_H = -\frac{g_{tt}}{g_{t\phi}}\bigg|_{r=r_2} = \frac{a}{r_2 r_0}. \tag{198}$$

Thence, the first law can be validated for the RLDBH through

$$dM_{QL} = T_H dS^{BH} + \Omega_H dJ. \tag{199}$$

Meanwhile, one may examine why Eq. (199) does not involve electric charge $Q$. This is because here $Q$ is a background charge of fixed value [125].

## 4.2 Quantization of the RLDBH with BQNMs

In the present section, we focus on the analysis of the entropy/area spectra of the RLDBHs [52]. Under scalar perturbations, we show that the radial equation reduces to a hypergeometric differential equation and imposing the QNM, NBC and DBC boundary conditions, we compute the BQNMs. Particularly, we derive the quantum entropy/area spectra of the RLDBH.

### 4.2.1 Separation of the Massless KGE on the RLDBH

Utilizing the massless KGE (19) with the following ansatz for the scalar field $\Psi$

$$\Psi = \frac{\rho(r)}{\sqrt{r}} e^{-i\omega t} Y_l^m(\theta, \phi), \quad Re(\omega) > 0, \tag{200}$$

where $Y_l^m(\theta, \phi)$ denotes the spheroidal harmonics with the eigenvalue $-l(l+1)$ [161], we get

$$\left\{Z\partial_r^2 + \left(\frac{r^2 - r_1 r_2}{r}\right)\partial_r + \left[\frac{(\omega r r_0 - ma)^2}{Z} - l(l+1) + \frac{(3r_1 - r)r_2 - rr_1 - r^2}{4r^2}\right]\right\}\rho(r) = 0, \tag{201}$$

From Eq. (201), one can obtain the ZE [61]

$$(-\partial_{r^*}^2 + V)\rho(r^*) = \omega^2 \rho(r^*), \tag{202}$$

where the effective potential is given by

$$V = f(r)\left\{\frac{1}{h(r)}\left[\frac{r^2+2Mr-3a^2}{4r^2} + l(l+1)\right] - \frac{m\tilde{\Omega}}{f(r)}(m\tilde{\Omega} - 2\omega)\right\}, \tag{203}$$

in which

$$\tilde{\Omega} = \frac{a}{rr_0}. \tag{204}$$

Defining the tortoise coordinate as $r^* = \int \frac{dr}{f(r)}$ yields

$$r^* = \frac{r_0}{r_2-r_1}ln\left[\frac{\left(\frac{r}{r_2}-1\right)^{r_2}}{(r-r_1)^{r_1}}\right]. \tag{205}$$

The asymptotic limits of $r^*$ become

$$\lim_{r \to r_2} r^* = -\infty, \tag{206}$$

$$\lim_{r \to \infty} r^* = \infty. \tag{207}$$

### 4.2.2 BQNMFs and Entropy/Area Spectra of RLDBH

In this section, we aim to read the BQNMFs of the caged RLDBH. These frequencies can be analytically determined when the confining mirrors are placed in the vicinity of the event horizon. We impose the Hod's boundary condition that nothing is supposed to come out of the event horizon of the caged BH and to survive at the cage [107].

Setting

$$y = \frac{r-r_2}{r_2}, \tag{208}$$

the metric function $f(r)$ transforms into

$$f(r) \to f(y) = \frac{1}{r_0}\left(r_2 y - \frac{y}{1+y}r_1\right), \tag{209}$$

which has the following NH behavior:

$$f_{NH}(y) \cong \tau y + O(y^2), \tag{210}$$

wherein $\tau = \frac{r_2 - r_1}{r_0}$. Thus, the tortoise coordinate around the NH behaves as

$$r^* \cong \int \frac{r_2 dy}{f_{NH}(y)} \cong \frac{r_2}{\tau} \ln y = \frac{1}{2\kappa} \ln y, \tag{211}$$

so that we have

$$y = e^{2\kappa r^*}. \tag{212}$$

Substituting Eq. (208) into Eq. (203), one finds the NH Zerilli potential as follows

$$V_{NH}(y) = \alpha + Fy + O(y^2), \tag{213}$$

where the parameters of Eq. (213) are given by

$$\alpha = m\Omega_H(\widetilde{\omega} + \omega), \tag{214}$$

$$F = \frac{\kappa}{2r_0 r_2^2}\{[(2l+1)r_2]^2 + 2Mr_2 - 3a^2\} - 2m\Omega_H\widetilde{\omega}, \tag{215}$$

in which

$$\widetilde{\omega} = \omega - m\Omega_H. \tag{216}$$

Here $\widetilde{\omega}$ connotes the wave frequency measured by the observer rotating with the horizon [125]. Therefore, the ZE (202) around the NH becomes

$$[-\partial_{r^*}^2 + \alpha + Fe^{2\kappa r^*}]\rho(r^*) = \omega^2 \rho(r^*). \tag{217}$$

One can find that the general solution of Eq. (217) is given by

$$\rho(r^*) = D_1 J_{-i\overline{\omega}}(u) + D_2 Y_{-i\overline{\omega}}(u), \tag{218}$$

where $u = 2i\sqrt{\eta}e^{\kappa r^*}$ and $D_1$ and $D_2$ are constants. $J_{-i\overline{\omega}}(u)$ and $Y_{-i\overline{\omega}}(u)$ denote the Bessel functions [113] of the first and second kinds, respectively. The parameters of the Bessel functions are given by

$$\eta = \frac{Fr_2^2}{\tau^2}, \tag{219}$$

$$\overline{\omega} = \frac{\widetilde{\omega}}{\kappa}. \tag{220}$$

Using the limiting forms (42) and (43), we get the NH ($e^{\kappa r^*} \ll 1$) behavior of Eq. (218):

$$\rho \sim D_1 \frac{(i\sqrt{\eta})^{-i\overline{\omega}}}{\Gamma(1-i\overline{\omega})} e^{-i\widetilde{\omega}r^*} - D_2 \frac{1}{\pi} \Gamma(-i\overline{\omega})(i\sqrt{\eta})^{i\overline{\omega}} e^{i\widetilde{\omega}r^*}. \quad (221)$$

One can deduce that $D_1$ and $D_2$ are related to the amplitude of the ingoing and outgoing waves at NH, respectively. Now, one can impose the condition of having QNMs, which stipulates purely ingoing waves at the horizon. Hence, we pick $D_1 \neq 0$ and set $D_2 = 0$. Thus, the physical solution of Eq. (218) is given by

$$\rho(r^*) = D_1 J_{-i\overline{\omega}}(u) \quad \text{or} \quad \rho(y) = D_1 J_{-i\overline{\omega}}(2i\sqrt{\eta y}). \quad (222)$$

Following [106,107,112], we consider the DBC at the surface $y \equiv y_m = \frac{r_m - r_2}{r_2}$ of the confining cage:

$$\rho(y)|_{y=y_m} = J_{-i\overline{\omega}}(2i\sqrt{\eta y_m}) = 0. \quad (223)$$

Using the relation (48) we can rewrite the condition (223) as

$$tan(i\pi\overline{\omega}) = \frac{J_{i\overline{\omega}}(2i\sqrt{\eta y_m})}{Y_{i\overline{\omega}}(2i\sqrt{\eta y_m})}, \quad (224)$$

which is called the resonance condition [106,107]. According to Hod [107], the boundary of the confining cavity leading to the characteristic resonance spectra is close to the NH ($r_m \approx r_2$). Thence, $l_m \equiv \eta y_m \ll 1$.

With the aid of Eqs. (42) and (43), we approximate the resonance condition (224) to

$$tan(i\overline{\omega}\pi) \sim -\frac{\pi(i\sqrt{l_m})^{2i\overline{\omega}}}{\Gamma(i\overline{\omega}+1)\Gamma(i\overline{\omega})} = i\frac{\pi e^{-\pi\overline{\omega}}}{\overline{\omega}\Gamma^2(i\overline{\omega})} (l_m)^{i\overline{\omega}}. \quad (225)$$

Finally, one should also impose the NBC [107] on the solution (222):

$$\frac{d\rho(y)}{dy}\bigg|_{y=y_m} = 0, \quad (226)$$

which admits the following expression [113]

$$J_{-i\overline{\omega}-1}(2i\sqrt{l_m}) - J_{-i\overline{\omega}+1}(2i\sqrt{l_m}) = 0. \quad (227)$$

Substituting Eq. (227) into Eq. (54), we derive the resonance condition for the NBC:

$$tan(i\pi\overline{\omega}) = \frac{J_{i\overline{\omega}-1}(2i\sqrt{l_m})}{Y_{i\overline{\omega}+1}(2i\sqrt{l_m})} \left[ \frac{-1+\frac{J_{i\overline{\omega}+1}(2i\sqrt{l_m})}{J_{i\overline{\omega}-1}(2i\sqrt{l_m})}}{1-\frac{Y_{i\overline{\omega}-1}(2i\sqrt{l_m})}{Y_{i\overline{\omega}+1}(2i\sqrt{l_m})}} \right]. \tag{228}$$

From Eqs. (42) and (43), we get

$$\frac{Y_{i\overline{\omega}-1}(2i\sqrt{l_m})}{Y_{i\overline{\omega}+1}(2i\sqrt{l_m})} \equiv \frac{J_{i\overline{\omega}+1}(2i\sqrt{l_m})}{J_{i\overline{\omega}-1}(2i\sqrt{l_m})} \sim O(l_m), \tag{229}$$

Thus, we can rewrite the resonance condition (228) as

$$tan(i\pi\overline{\omega}) \sim -\frac{J_{i\overline{\omega}-1}(2i\sqrt{l_m})}{Y_{i\overline{\omega}+1}(2i\sqrt{l_m})} = -i\frac{\pi e^{-\pi\overline{\omega}}}{\overline{\omega}\Gamma^2(i\overline{\omega})}(l_m)^{i\overline{\omega}}. \tag{230}$$

Since $l_m \equiv \eta y_m \ll 1$, the damped modes with $Im(\overline{\omega}) < 0$ lead to $(l_m)^{i\overline{\omega}} \ll 1$. So, it is clear that both resonance conditions (225) and (230) take small quantities.

We now use an iteration scheme to obtain the BQNMs of the caged RLDBH. For both DBC and NBC, the $0^{th}$ order resonance condition is given by [107]:

$$tan(i\pi\overline{\omega}_n^{(0)}) = 0, \tag{231}$$

which results in

$$\overline{\omega}_n^{(0)} = -in, \quad n = 1, 2, 3, .... \tag{232}$$

The $1^{st}$ order resonance condition is obtained after substituting Eq. (232) into r.h.s. of Eqs. (225) and (230). Thus, we have

$$tan(i\overline{\omega}_n^{(1)}\pi) = \pm i \frac{\pi e^{i\pi n}}{(-in)\Gamma^2(n)}(l_m)^n, \tag{233}$$

which is equivalent to

$$tan(i\overline{\omega}_n^{(1)}\pi) = \mp n\frac{\pi(-l_m)^n}{(n!)^2}, \tag{234}$$

where plus (minus) stands for the NBC (DBC). In the $u \ll 1$ regime, we use Eq. (62) for Eq. (234) and find the characteristic resonance spectra as follows

$$i\overline{\omega}_n\pi = n\pi \mp n\frac{\pi(-l_m)^n}{(n!)^2}, \tag{235}$$

which yields

$$\overline{\omega}_n = -in\left[1 \mp \frac{(-l_m)^n}{(n!)^2}\right].$$ (236)

Putting this into Eqs. (220) and (216) gives the BQNMs of the RLDBH

$$\omega_n = m\Omega_H - i\kappa n\left[1 \mp \frac{(-l_m)^n}{(n!)^2}\right], \quad n = 1, 2, 3, \ldots$$ (237)

For the highly damped modes ($n \to \infty$), Eq. (237) becomes

$$\omega_n \approx m\Omega_H - i\kappa n,$$ (238)

which is in accordance with the recent study of Sakalli [50] in which the quantization of the RLDBH was studied with the standard QNMs. According to MM, the transition frequency now reads

$$\Delta\omega \approx Im(\Delta\omega) = Im(\omega_n - \omega_{n-1}) = \kappa = \frac{2\pi T_H}{\hbar}.$$ (239)

From Eqs. (3) and (239), we now have

$$I_{adb} = \frac{\hbar}{2\pi} S^{BH}.$$ (240)

Acting upon the BSQR [27], we obtain the entropy spectrum as

$$S_n^{BH} = 2\pi n,$$ (241)

which also yields the area spectrum as follows

$$\mathcal{A}_n = 4\hbar S_n^{BH} = 8\pi\hbar n.$$ (242)

Thus, the minimum area spacing of the RLDBH becomes

$$\Delta\mathcal{A}_{min} = 8\pi\hbar.$$ (243)

The above results support Bekenstein's conjecture [8]. Furthermore, we see that the spectroscopy of the RLDBH is independent from its characteristic parameters, and its area spectrum is evenly spaced.

## 4.3 Cloud Heights of the MRLDBH

In this section, we consider a charged massive scalar field coupled to a RLDBH. For a comprehensive analytical study, the RLDBH is assumed to be extremal (i.e.,

MRLDBH), which requires the equality of mass term ($M$) with the rotation term ($a$). Namely, we focus on the case of a charged massive test scalar field in the geometry of the MRLDBH spacetime [163].

### 4.3.1 Separation of the KGFE on the MRLDBH

The dynamics of a charged massive scalar field $\Psi$ in the RLDBH spacetime is governed by the KGFE (see, for example, [164]):

$$\left(\partial_\mu - iqA_\mu^{em}\right)\left(\sqrt{-g}g^{\mu\nu}(\partial_\nu - iqA_\nu^{em})\Psi\right) - \sqrt{-g}\mu^2\Psi = 0, \quad (244)$$

where $q$ and $\mu$ are the charge and mass of the scalar particle, respectively. We assume the ansatz for $\Psi$, as follows:

$$\Psi \equiv \Psi_{lm}(t,r,\theta,\phi) = e^{im\phi}S_{lm}(\theta)R_{lm}(r)e^{-i\omega t}, \quad (245)$$

where $\omega$ is the conserved frequency of the mode, and $l$ and $m$ are the spheroidal and azimuthal harmonic indices, respectively, with $-l \leq m \leq l$. In Eq. (245), $R_{lm}$ and $S_{lm}$ are the functions of radial and angular equations of the confluent Heun differential equation with the separation constant $\lambda_{lm}$ [123,165-167].

For the MRLDBH ($M = a$), the angular part of Eq. (244) obeys the following differential equation of spheroidal harmonics $S_{lm}$ [113,165-169]:

$$\frac{1}{\sin\theta}\frac{\partial}{\partial\theta}\left(\sin\theta\frac{\partial S_{lm}}{\partial\theta}\right) + \left[\lambda_{lm} - \left(\frac{1}{2}\hat{q}M\sin\theta\right)^2 - \frac{m^2}{\sin^2\theta}\right]S_{lm} = 0, \quad (246)$$

where $\hat{q} = \sqrt{2}q$.

The above differential equation has two poles at $\theta = 0$ and $\theta = \pi$. For a physical solution, $S_{lm}(\theta)$ functions are required to be regular at those poles. This remark enables us to obtain a discrete set of eigenvalues $\lambda_{lm}$. In Eq. (246), $\frac{1}{2}M^2q^2\cos^2\theta$ can be treated as a perturbation term on the generalized Legendre equation [113]. Thus, we obtain the following perturbation expansion:

$$\lambda_{lm} - \left(\frac{1}{2}\hat{q}M\right)^2 = \sum_{k=0}^{\infty} c_k(-1)^k \left(\frac{1}{2}\hat{q}M\right)^{2k}. \tag{247}$$

It is worth noting that in [113], the expansion coefficients in the summation symbol of Eq. (247), $\{c_k(l,m)\}$, are explicitly given.

The radial part of the KGFE (244) in the MRLDBH geometry acts as a Teukolsky equation [168,169]:

$$\Delta^2 \frac{d}{dr}\left(\Delta^2 \frac{dR_{lm}}{dr}\right) + \{\mathcal{H}^2 - mM[2\hat{q}Mr - mM + 2h(r)\omega]\}R_{lm}$$

$$-\Delta^2[K + h(r)\mu^2]R_{lm} = 0, \tag{248}$$

where

$$\Delta = r - M, \tag{249}$$

and

$$\mathcal{H} = h(r)\omega + \frac{\hat{q}}{2}(r^2 + M^2). \tag{250}$$

For the MRLDBH, the event horizon is located at $r_{EH} = M$, which is the degenerate zero of Eq. (249).

In general (for AF BHs), the bound nature of the scalar clouds obeys the following boundary conditions: purely ingoing waves at the event horizon and decaying waves at spatial infinity [64,170-175]. However, when we study the limiting behaviors of the radial equation (248), we get

$$R_{lm} \sim \begin{cases} \frac{1}{\sqrt{r}} J\left(0, qe^{\frac{r^*}{r_0}}\right) \approx \frac{1}{r}\sqrt{\frac{2}{\pi q}} \sin\left(qe^{\frac{r^*}{r_0}} + \frac{\pi}{4}\right) & \text{as } r \to \infty \ (r^* \to \infty) \\ \frac{1}{\sqrt{r_{EH}}} e^{-i[\omega - (m-\hat{q}M)\widetilde{\Omega}]r^*} & \text{as } r \to r_{EH} \ (r^* \to -\infty), \end{cases} \tag{251}$$

where $\tilde{\Omega} = (r_0)^{-1}$ is the angular velocity of $r_{EH}$, and $r^*$ is the tortoise coordinate of the MRLDBH:

$$r^* = \int \frac{rr_0 dr}{\Delta^2} = r_0 \ln(\Delta) - \frac{Mr_0}{\Delta}. \tag{252}$$

It is clear from the asymptotic solution (251) that unlike the Kerr BH [64], there are no decaying (bounded) waves in MLRDBH geometry at spatial infinity. Instead of this, we have oscillatory but fading waves (because of the factor $\frac{1}{r}$) at the asymptotic region (see Figure 3). This result probably originates from the NAF structure of the MRLDBH geometry.

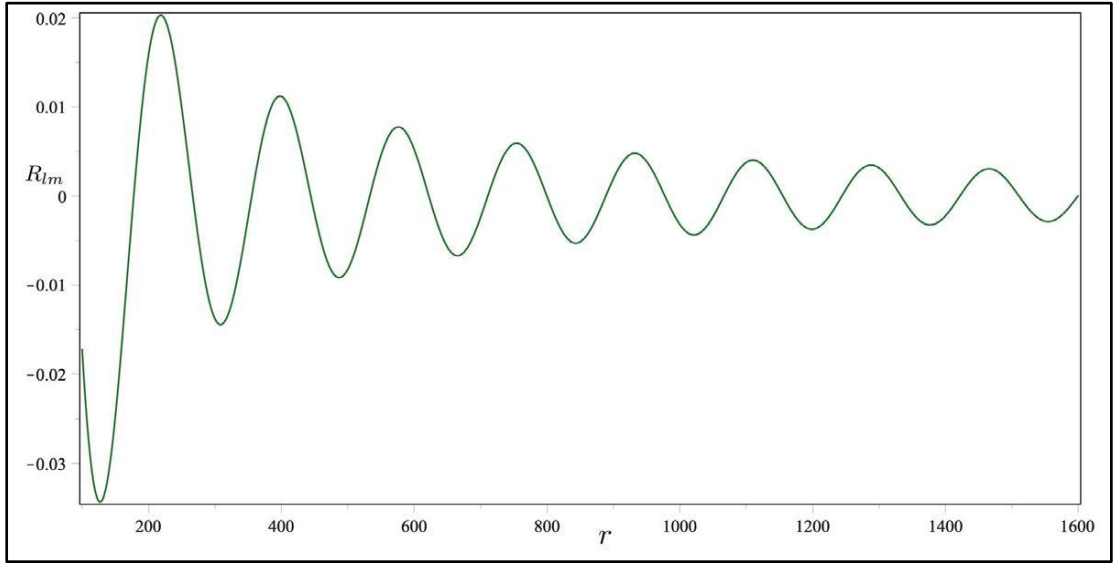

Figure 3: Asymptotic behavior of radial function $R_{lm}(r)$. The plot is governed by Eq. (251). The physical parameters are chosen as $M = 1$, $q = 0.05$, and $r_0 = 0.5$.

On the other hand, a very recent study [175] has shown that scalar clouds can have a semipermeable surface, and thus they might serve as a `partial confinement' in which only outgoing waves are allowed to survive at the spatial infinity. Considering this fact, we will impose an asymptotic boundary condition to those non-decaying (unbounded) scalar clouds: only pure outgoing waves propagate at spatial infinity.

### 4.3.2 Stationary Resonances and Effective Heights of the Scalar Clouds Around the MRLDBH

Stationary resonances or the so-called marginally stable modes are the stationary regular solutions of the KGFE (244) around the horizon. They are characterized by $Im\,\omega = 0$ [64], which corresponds to $\omega = (m - \hat{q}M)\widetilde{\Omega}$ for the MRLDBH spacetime. In fact, such resonances saturate the superradiance condition [125].

Introducing a new dimensionless variable

$$x = \frac{\Delta}{M}. \tag{253}$$

one can see that the radial equation (248) can be rewritten as

$$x^2 \frac{d^2 R_{lm}}{dr^2} + 2x \frac{dR_{lm}}{dr} + V_{eff} R_{lm} = 0, \tag{254}$$

in which the effective potential is given by

$$V_{eff} = \left(\frac{1}{2}\hat{q}Mx\right)^2 + M(1+x)(m\hat{q} - r_0\mu^2) + m^2 - \lambda_{lm}. \tag{255}$$

Letting

$$Y = xR_{lm} \quad \text{and} \quad z = -i\hat{q}Mx, \tag{256}$$

Eq. (254) transforms into the following differential equation

$$\frac{d^2Y}{dz^2} + \left(-\frac{1}{4} + \frac{\sigma}{z} + \frac{\frac{1}{4}-\beta^2}{z^2}\right)Y = 0, \tag{257}$$

with

$$\sigma = i\left(m - \frac{r_0}{\hat{q}}\mu^2\right) \quad \text{and} \quad \beta^2 = \lambda_{lm} + \frac{1}{4} - m^2 + (r_0\mu^2 - m\hat{q})M. \tag{258}$$

Without loss of generality, one can assume that $\beta$ is a non-negative real number [64]. Therefore, Eq. (257) corresponds to a Whittaker equation [113], whose solutions can be expressed in terms of the CH functions $M(a, b, z)$ [113,143]. Thus, the solution of Eq. (254) can be given by

$$R_{lm} = \frac{e^{-\frac{1}{2}z}}{\sqrt{z}}\left[C_1 z^\beta M\left(\frac{1}{2}+\beta-\sigma, 1+2\beta, z\right) + C_2 z^{-\beta} M\left(\frac{1}{2}-\beta-\sigma, 1-2\beta, z\right)\right], \quad (259)$$

where $C_1$ and $C_2$ are the integration constants. Near the horizon, solution (259) reduces to [177]

$$R_{lm} \to C_1 z^{-\frac{1}{2}+\beta} + C_2 z^{-\frac{1}{2}-\beta}. \quad (260)$$

Since the NH solution $(z \to 0)$ must admit the regularity, one can figure out that

$$C_2 = 0 \text{ and } \beta \geq \frac{1}{2}. \quad (261)$$

By using the asymptotic behaviors of the CH functions [113,143], Eq. (259) can be approximated (for $z \to \infty$) to

$$R_{lm} \to C_1\left[e^{\frac{1}{2}z} \frac{\Gamma(1+2\beta)}{\Gamma(\frac{1}{2}+\beta-\sigma)} z^{-1-\sigma} + e^{-\frac{1}{2}z} \frac{\Gamma(1+2\beta)}{\Gamma(\frac{1}{2}+\beta+\sigma)} z^{-1+\sigma}(-1)^{-\frac{1}{2}-\beta+\sigma}\right]. \quad (262)$$

Recalling the complex structure of $z$ (see Eq. (256)), we infer that the first term in the square bracket of Eq. (262) stands for the asymptotic ingoing waves, however the second one represents the asymptotic outgoing waves. According to the physical boundary conditions aforementioned, the asymptotic ingoing wave ($\sim e^{\frac{1}{2}z}$) in Eq. (262) must be terminated. This is possible by employing the pole structure of the Gamma function ($\Gamma(\tau)$ has the poles at $\tau = -n$ for $n = 0, 1, 2, ...$ [113].). Therefore, the resonance condition for the stationary unbound states of the field eventually becomes

$$\frac{1}{2} + \beta - \sigma = -n. \quad (263)$$

It is convenient to express the radial solution of the unbound states in a more compact form by the generalized Laguerre polynomials $L_n^{(2\beta)}(z)$ [113]:

$$R_{lm} = C_1 z^{-\frac{1}{2}+\beta} e^{-\frac{1}{2}z} L_n^{(2\beta)}(z). \quad (264)$$

One can deduce from the resonance condition (263) that $\sigma$ should be a real number. However, taking cognizance of Eq. (258), which indicates that $\sigma$ is a pure imaginary parameter, we conclude that the resonances correspond to $\sigma = 0$. Thus, we have two cases:

*Case-I*:

$$m = \mu = 0, \qquad (265)$$

and

*Case-II*:

$$\hat{q} = \frac{r_0 \mu^2}{m}. \qquad (266)$$

It is worth noting that both cases exclude the existence of regular static ($\omega = 0$) solutions since $\omega = (m - \hat{q}M)\widetilde{\Omega}$. The latter remark is in accordance with the famous no-hair theorems [7,63,68,71-74,97,98], since they exempt the static hairy configurations [65].

For solving the resonance condition (263), it is practical to introduce another dimensionless variable:

$$\epsilon = \frac{i}{2}\hat{q}M, \qquad (267)$$

so that Eq. (258) can be rewritten as [64,113]

$$\sigma = \frac{\sqrt{2}M\mu^2 Q + 2im\varepsilon}{2\varepsilon}, \qquad (268)$$

$$\beta^2 = \left(l + \frac{1}{2}\right)^2 - m^2 - \varepsilon^2 + M\mu^2 r_0 + 2im\varepsilon + \sum_{k=1}^{\infty} c_k \varepsilon^{2k}. \qquad (269)$$

After substituting Eqs. (268) and (269) into Eq. (263), we express the resonance condition as a polynomial equation for $\varepsilon$ :

$$8[l(l+1) + m^2 - 1]\varepsilon^4$$
$$+(2l-1)(2l+3)\{-8im\varepsilon^3 - 4[im(2n+1) - n(n+1) + l(l+1) + r_0 M\mu^2]\varepsilon^2$$
$$+2r_0 M\mu^2[2im - (2n+1)]\varepsilon + (r_0 M\mu^2)^2 - 4\sum_{k=2}^{\infty} c_k \varepsilon^{2k+2}\} = 0. \qquad (270)$$

Discrete and infinite group of stationary resonances for both cases are presented in Table 1 and Table 2. The results are shown for the different values of $n$ (resonance parameter). Unlike the Kerr BH [64], our numerical calculations about Eq. (270) showed that in the case of $n = 0$ the obtained resonance values ($Mq_\text{resonance}$) are complex, which do not admit physically acceptable results. For this reason, we consider the resonance parameters of having $n \geq 1$.

Table 1: Stationary scalar resonances and cloud heights of a MRLDBH for *Case-I*. $Mq_\text{resonance}$ and $|x|_\text{cloud}^{(n)}$ are tabulated for an *s-wave* ($l = m = 0$) with $n \geq 1$.

| $n$ | $Mq_\text{resonance}$ | $|x|_\text{cloud}^{(n)}$ |
|---|---|---|
| 1 | 2.4495 | 1.7321 |
| 2 | 4.2426 | 1.6667 |
| 3 | 6.0000 | 1.6499 |
| 4 | 7.7460 | 1.6432 |

Table 2: Stationary scalar resonances and cloud heights of a MRLDBH for *Case-II*. $Mq_\text{resonance}$ and $|x|_\text{cloud}^{(n)}$ are represented for the fundamental resonances $l = m = 1$ with $n \geq 1$. Qualitatively similar results are observed for other values of $\{l, m\}$.

| $n$ | $Mq_\text{resonance}$ | $|x|_\text{cloud}^{(n)}$ |
|---|---|---|
| 1 | 1.5811 | 2.6833 |
| 2 | 3.5355 | 2.0000 |
| 3 | 5.2440 | 1.8878 |
| 4 | 6.8920 | 1.8468 |

We now consider the effective heights of the stationary charged massive scalar field configurations surrounding the MRLDBH. These configurations correspond to the group of wave functions (264) that fulfill the resonance condition (263). According to the no-short hair theorem proposed for the spherically symmetric and static hairy BH configurations [19], the hairosphere [177] must extend beyond $\frac{3}{2} r_{EH}$. Taking cognizance of Eq. (253), we conclude that the minimum radius of the hairosphere corresponds to $|x|_{\text{hair}} = \frac{1}{2}$, where $|x|$ is the dimensionless height (absolute altitude). Furthermore, we can compute the size of the stationary scalar clouds by defining their effective radii. The effective heights of the scalar clouds can be approximated to a radial position at which the quantity $4|x|^2|\Psi|^2$ reaches its global maximum value [64]. By using Eq. (264), one finds the dimensionless heights of the clouds, as follows:

$$|x|_{\text{cloud}}^{(n)} = \left|\frac{2\beta+1+2n}{2\varepsilon}\right|, \quad n = 1, 2, 3, \dots \quad (271)$$

Note that, according to the private communications between Hod (the author of [174]) and us, it is understood that there is a typo in the expression of 'effective radii as the radii at which the quantity $4r^2|\Psi|^2$ attains its global maximum' of [174]: In the expression of $4r^2|\Psi|^2$, $r$ should be replaced with the dimensionless height $|x|$.

The effective heights of the principal clouds above the central BH for both *Case-I* and *Case-II* are also displayed in Table 1 and Table 2, respectively. It can be easily seen from Table 1 and Table 2, $\left\{|x|_{\text{cloud}}^{(n)}\right\}$ are always larger than the lower bound of the hairosphere ($|x|_{\text{hair}} \geq \frac{1}{2}$) [64].

# Chapter 5

# CONCLUSION

In this thesis, the quantum entropy/area spectra of the GHSBH are first investigated from two different approaches (the BQNMs and the RMs) [106,149] that are based on the adiabatic invariant formulation of BHs in MM. For this purpose, we have first separated the massless KGE into the angular and the radial parts on the GHSBH geometry and showed that the SLE (or ZE) with its effective potential can be obtained from the radial equation. The Zerilli potential is obtained in the NH region.

Motivated by the recent studies [106,107] which have shown that the BQNMs are formed when a BH is confined by a perfect mirror in the NH region, we first attempted to find the BQNMs (resonance spectra) of the GHSBH. To this end, we considered that the GHSBH is confined by a mirror which is located very close to the BH horizon ($r_m \approx r_+$) [107]. The ZE is perfectly approximated by a Bessel differential equation in NH region. After imposing the QNM boundary conditions with the DBC and NBC, the BQNMFs of the caged GHSBH are obtained. Then MM is applied to the BQNMFs for the ultrahigh damping, and the entropy/area spectra of the GHSBH are derived to be equidistant and independent of the BH parameters as stated by Wei et al. [25] and Kothawala et al. [42] who stated that the BHs in Einstein's gravity theory have equispaced area spectrum. The result obtained from the BQNMs of the GHSBH is same as the original result of Bekenstein: $\varepsilon = 8\pi$.

In our second attempt we read the RMs of the GHSBH for $l \gg 1$. The effective potential has a BP just beyond the event horizon, as seen from the Figure 1, causing the scalar waves to be trapped between the BP and the horizon. Then, in order to find the RMs in the NH region, a particular approximation method [45,48,51,114] is applied. The one-dimensional SLE is well approximated to a CH differential equation. It is straightforward to compute the RMs of the GHSBH and in sequel to apply MM for the highly damped RMFs. Doing so, we found similar results to the previous one (i.e., equidistant spectra which are independent of the physical parameters of the BH), however our calculations revealed that the value of the dimensionless constant is now $\varepsilon = 16\pi$ and is questionable. Hod [37] underlined that the uniform quantization of the BH spectra is more important than the value of $\varepsilon$.

In Chapter 3, we studied the quantum spectra of the ZZLBH in the presence of two different fields. By using the MM, we obtained the QNMs of the ZZLBH [155] as well as the Dirac BQNMs [157].

For obtaining the QNMs, a SLE with its associated potential is found after separating the massive KGE on the ZZLBH background. The behaviors of the potential around the horizon and at the spatial infinity are checked, and as depicted in Figure 2, the potential never dies off at the spatial infinity. The obtained SLE is approximated to a CH differential equation and the QNMs of the ZZLBH are derived with the aid of the pole structure of the Gamma functions which appear in the solution. For $n \to \infty$, we applied MM to the QNMFs and calculated the entropy/area spectra. Although the obtained spectra are both equispaced and free from the physical parameters of the ZZLBH, as expected; we obtained a different equispacing in the area with $\varepsilon = 16\pi$ than the usual value [16,23,24]). However, Hod [37] showed that the area spacing (hence $\varepsilon$) may be different depending on the method applied for exploring the quantization of BHs. Moreover, our findings agree with the findings of Bekenstein [10] as well as the studies of Wei et al. [25] and Kothawala et al. [42].

Our next target was to study the Dirac QNMs of the ZZLBH to investigate the spin effect on the quantization. For this purpose, the Dirac equation for the uncharged massless spin-1/2 test particles on the ZZLBH background is decoupled with an eigenvalue $\lambda$ into radial and the angular equations, and then, the ZEs with their potentials are limited to the NH region for finding analytically the possible Dirac BQNMs in the presence of a fermionic field on this background by using the iteration scheme that is mentioned in Chapter 2. For this reason, we considered that the ZZLBH is caged by a mirror in the NH region and read the Dirac BQNMs whose frequencies are negative and purely imaginary which assures that the ZZLBHs are stable under massless fermionic field perturbations. After applying the appropriate boundary conditions, we computed the Dirac BQNMFs of the caged ZZLBH. And then imposing MM, the Dirac BQNMFs for ultrahigh dampings are used to derive the ZZLBH's entropy/area spectra which is shown to be equidistant and independent of the BH parameters as stated in [6-10,25,42], although with the Dirac field it is found that $\varepsilon = 4\pi$ which is different than Bekenstein's original result [16,23,24] and the results obtained from the QNM studies of [118,155].

In Chapter 4, firstly we computed the BQNMs of the RLDBH by studying the dynamics of the massless scalar field in the RLDBH background and explored the entropy/area spectra of RLDBHs from its BQNMs [52]. Secondly, we analyzed the dynamics of a charged massive scalar field in the background of MRLDBH [163]. We showed the existence of an infinite quantized set of resonances describing non-decaying charged massive scalar clouds around the MRLDBH and analytically calculated the effective heights of these clouds above the horizon of the MRLDBH.

For studying the quantization of the RLDBH, we first used the separability of the massless KGE to derive the ZE in this geometry and extracted its effective potential. Then, we showed how the solution of the NH ZE is expressed in terms of the Bessel functions. In this way, we computed the BQNMs analytically, after imposing the QNM boundary condition (only ingoing waves are allowed at the NH), DBC, and NBC. Then, MM is applied to find the transition frequencies from the highly damped BQNMs to derive the entropy/area spectra of the RLDBH. The resulting area spectrum (where $\varepsilon = 8\pi$) is equidistant and fully consistent with Bekenstein's original result [8] and Sakalli's result [50], which was obtained from the ordinary QNMs.

During the analysis of the charged massive scalar field dynamics in the MRLDBH, we solved the KGFE in this geometry. Separation of this equation revealed that the radial and the angular equations are of the confluent Heun differential equation with a separation constant and the angular part obeys the differential equation of spheroidal harmonics. Using the regularity property of the spheroidal harmonics at its poles and approximating this differential equation to a generalized Legendre equation, we obtained a discrete set of eigenvalues which can be expressed as a perturbation expansion. And, the radial equation is nothing but the Teukolsky equation. Asymptotic behaviors of the radial solution showed that instead of decaying (bounded) waves, we have fading oscillatory waves which can be seen from Figure 3. These can be considered as non-decaying (unbounded) scalar clouds at the asymptotic region of the MRLDBH. Later, the radial equation is transformed into a Whittaker equation whose solutions are given in terms of the CH functions and by using the asymptotic behaviors of the CH functions, the radial solution of the unbounded states is expressed in terms of the generalized Laguerre polynomials. By imposing the asymptotic boundary condition, we derived a resonance condition and concluded that there are two possible cases for having stationary scalar resonances. The fundamental resonances $l = m = 0$ and $l = m = 1$ with $n \geq 1$ are illustrated in Table 1 and Table 2, respectively. However, in both cases the resonance spectrum does not hold for the ground state, $n = 0$. At this point it is convenient to point out that our findings give support to the existence of non-decaying scalar field dark matter halo around the rotating BHs. It is also worth noting that the other combinations of $\{l, m\}$-values give almost the same results. In addition, we analytically computed the effective heights of the scalar clouds and showed that $|x|_{\text{hair}} \geq \frac{1}{2}$. At this juncture, one can interrogate our findings about whether they are compatible with no-short hair theorem [177] or not. Because, in the seminal works of Hod [145,146], it was argued that the charged rotating BHs may evince the failure of the no-short hair theorem since they have short bristles, however, the most recent work of Hod [146] has supported the no-short hair theorem. RLDBHs indeed show similarity to the Kerr BH due to having a fixed background charge which does not exist in the horizons but tunes the radius of the spherical part of its metric. They have no zero-charge limit [125]. Thus, similar to [64,146], our results give also support to the no-short hair theorem.

Extremal BHs are believed to have connection with the ground states of the QGT, which remarks the significance of the spin-1/2 particles on such backgrounds. Therefore, it would be interesting to extend these studies to analyze the dynamics of a Dirac field (a charged massive field having spins other than zero) interacting with a rotating (might be charged) BH.